\documentclass[acmsmall]{acmart}
\AtBeginDocument{%
  }

\setcopyright{acmlicensed}
\copyrightyear{2026}
\acmYear{2026}
\acmDOI{XXXXXXX.XXXXXXX}

\acmSubmissionID{TSC-2026-0016}

\begin{document}

\title{Diagnosing Urban Street Vitality: A Visual-Semantic and Spatiotemporal Framework for Street-Level Economics}

\author{Xinxin Zhuo}
\email{zhuoxinxin@seu.edu.cn}
\orcid{0009-0001-6642-5357}
\affiliation{%
  \institution{School of Economics and Management, Southeast University}
  \city{Nanjing}
  \country{China}
}
\author{Mengyuan Niu}
\orcid{0009-0009-9269-9560}
\email{220230918@seu.edu.cn}
\affiliation{%
  \institution{School of Information Science and Engineering, Southeast University}
  \city{Nanjing}
  \country{China}
}

\author{Ruizhe Wang}
\orcid{0009-0001-2511-8692}
\email{wangjh0521@seu.edu.cn}
\affiliation{%
  \institution{School of Information Science and Engineering, Southeast University}
  \city{Nanjing}
  \country{China}
}

\author{Junyan Yang}
\orcid{0000-0003-3895-4689}
\email{yjy-2@163.com}
\affiliation{%
  \institution{School of Architecture, Southeast University}
  \city{Nanjing}
  \country{China}
}
\author{Qiao Wang}
\authornote{Corresponding author.} 
\email{qiao.wang@seu.edu.cn}
\orcid{0000-0002-5271-0472}
\affiliation{%
  \institution{School of Economics and Management, Southeast University}
  \city{Nanjing}
  \country{China}
}
\affiliation{%
  \institution{School of Information Science and Engineering, Southeast University}
  \city{Nanjing}
  \country{China}
}

\renewcommand{\shortauthors}{Zhuo et al.}

\begin{abstract}
Micro-scale street-level economic assessment is fundamental for precision spatial resource allocation. While Street View Imagery (SVI) advances urban sensing, current approaches remain semantically superficial, neglecting brand hierarchy heterogeneity and structural recession. To decode this landscape, we introduce a Visual-Semantic and Field-Based Spatiotemporal Framework, operationalized via the Street Economic Vitality Index (SEVI).

Our approach synergizes physical and semantic streetscape parsing. We perform instance segmentation to identify active signboards, glass interfaces, and storefront closures. A dual-stage VLM-LLM pipeline processes active signage, standardizing variants into global hierarchies to quantify a spatially smoothed brand premium index. To overcome static SVI limitations and introduce causal awareness, we implement a temporal lag design using subsequent dynamic Location-Based Services (LBS) data to represent realized spatial demand. Coupling this with a category-weighted Gaussian spillover model—shifting from rigid buffers to continuous fields—we establish a tri-dimensional diagnostic system (Commercial Activity, Spatial Utilization, Physical Environment).

Empirical results from a time-lagged Time-Sliced Geographically Weighted Regression across eight tidal periods in Nanjing reveal quasi-causal mechanisms of spatiotemporal heterogeneity. Street vibrancy emerges from a time-dependent synergy between historical hierarchical brand clustering and mall-induced spatial externalities. High-quality commercial interfaces exhibit quasi-causal attractive power peaking during discretionary midday and evening hours, while structural recession (closures) exerts a severe, lagged night-time repulsion effect. Extensive robustness checks confirm the absolute structural stability of these findings. Ultimately, this framework provides actionable, evidence-based strategies for precision spatial governance.
\end{abstract}





\maketitle

\section{Introduction}

A persistent puzzle in urban economics is why streets with seemingly identical physical infrastructure, geographical advantages, and high establishment densities often exhibit drastically divergent economic trajectories. Urban policymakers and investors increasingly confront the "illusion of prosperity"—situations where high nominal commercial density masks underlying vulnerability, low economic yield, and hidden recession. Understanding and resolving this divergence is fundamental to deciphering the microscale dynamics of urban life and correcting the misallocation of spatial resources during urban renewal \cite{Li2022_IJGI, Ying2019, He2024, Yang2024, Chen2022}.

Traditional urban economic models and conventional metrics struggle to explain this micro-level divergence. Metrics aggregated at district or grid levels often obscure the fine-grained interactions among commercial establishments and pedestrian flows \cite{KOO2023, Chen2019, Zhang2021_EPB}. While the advent of big data has introduced Point of Interest (POI) density as a proxy for vitality, this approach suffers from critical information asymmetry. It frequently identifies ``zombie stores'' (establishments that remain registered but are operationally closed) as active participants, thereby yielding false positives and failing to capture actual market exit behaviors. To address this data-driven opacity, recent scholarship has employed Street View Imagery (SVI) \cite{Jiang2022, Li2022, Zhang2019_SocialSensing} to visually audit store closures. However, these visual approaches remain reductionist, treating all active storefronts as homogenous entities based on a binary ``open vs. closed'' status.

Consequently, existing literature overlooks two decisive micro-mechanisms driving this economic divergence: the quality heterogeneity of commercial presence (e.g., the economic disparity between a local survival-oriented shop and a global premium chain) and the spatial externalities of large commercial anchors. This semantic opacity prevents a true understanding of whether a street's vitality stems from endogenous high-quality commercial agglomeration or exogenous reliance on a nearby commercial anchor. Furthermore, the inherent complexity of storefront signage represents unstructured market signals that traditional Optical Character Recognition (OCR) fails to decode, leaving the brand-driven spillover effects unquantified.

To untangle this economic puzzle, we propose a visual-semantic and spatiotemporal framework for assessing street-level economic vitality, moving beyond simple enumeration toward a diagnostic evaluation of urban commercial health. We construct a Street Economic Vitality Index (SEVI) and demonstrate its applicability through an empirical study of central Nanjing. We conceptualize street vitality as the result of a dynamic synergy between internal commercial quality and external radiative fields. Specifically, we leverage Vision-Language Models (VLMs) coupled with Large Language Model (LLM) reasoning to extract and refine brand semantics from signboards. This allows for the precise calculation of a Weighted Brand Ratio Index, which differentiates between local and international commercial tiers, effectively pricing the quality heterogeneity of street-level retail. 

Crucially, we extend this framework by rigorously modeling Mall Spillover Vitality ($MV_i$) to capture the spatial externalities of large commercial anchors. While foundational studies have significantly advanced commercial agglomeration analysis \cite{Sevtsuk2014, Yue2017}, they typically operationalize influence through uniform gravity metrics. These approaches homogenize the spatial externalities of diverse commercial entities. By calibrating specific Gaussian decay rates ($\sigma$) contingent on semantic categorization within a defined spatial boundary (e.g., a 2,000-meter localized threshold), we shift from threshold-based to field-based assessment. This dual-model approach enables us to distinguish areas where vitality is driven by structural advantages (field energy) versus those suffering from hidden recession. Furthermore, to overcome the temporal limitations of static visual audits, we incorporate dynamic Location-Based Services (LBS) data to capture the true tidal rhythms of street-level human activity.

Nanjing’s central districts, designated as the primary development axis in the \textit{Territorial and Spatial Master Plan (2021--2035)} \cite{NanjingPlan2024}, serve as an ideal case study. The area faces the dual challenge of aging infrastructure and fragmented commercial fabrics. By operationalizing SEVI, we provide a tool that not only measures current vitality but also unveils the latent mechanisms of quality-driven growth and spatial energy essential for precision spatial governance.

This study advances the discourse on urban economic vitality through four main contributions:
\begin{enumerate}
    \item \textbf{From Binary Detection to Quality Heterogeneity}: Transcending simple occupancy checks, we perform a deep semantic assessment of commercial quality. By leveraging VLM-driven recognition and LLM-based rectification, the framework achieves high-fidelity standardization of brand data, effectively converting visual signals into weighted economic indicators that reflect market-tier distinctions.
    
    \item \textbf{Field-Based Modeling of Spatial Externalities}: We introduce a Gaussian decay function to quantify the continuous spatial attenuation of Mall Spillover Vitality ($MV_i$). This shifts the analysis from traditional buffer-based metrics to a field-based assessment, enabling a more nuanced representation of how diverse commercial anchors exert spatial spillover effects over their surrounding urban fabric.
    
    \item \textbf{Spatiotemporal Decoupling via Temporal Lag Design}: To avoid simultaneity bias and move beyond purely descriptive co-occurrences, we synergize high-resolution Location-Based Services (LBS) data with a time-lagged Time-Sliced Geographically Weighted Regression (GWR). By treating the multi-source visual and semantic metrics as a historical physical baseline to predict subsequent dynamic crowd intensity, our framework successfully identifies the quasi-causal drivers of street vitality and reveals how their attractive power fluctuates between rigid (commuting) and discretionary (leisure) travel purposes.
    
    \item \textbf{Actionable Framework for Precision Spatial Governance}: Beyond theoretical measurement, we operationalize these metrics into a diagnostic tool (SEVI). This allows policymakers to distinguish between areas of structural advantage (driven by field energy) and hidden recession (masked by high nominal density), providing evidence-based guidance for the optimal allocation of spatial resources.
\end{enumerate}

\section{Related Work} 
Research on urban economic vitality has undergone a paradigm shift from macroscopic statistical analysis to microscopic street-level sensing. In this section, we review the evolution of these measurement methodologies, specifically critiquing current limitations regarding semantic depth (commercial heterogeneity) and spatial modeling (agglomeration externalities), which motivate our methodological interventions.

\subsection{From Macro-Statistics to Micro-Sensing} 
Early assessments of urban vitality relied heavily on macro-level socio-economic indicators, such as regional GDP and employment density \cite{Wang2022_JUM}. Although effective for regional comparisons, these metrics lack the granular resolution required for community-level spatial governance. The advent of Big Data introduced meso-scale proxies, such as POI density and LBS positioning data \cite{Lan2020, Chen2019}. However, these datasets are inherently prone to an observational lag in market exit—they record registered entities but fail to capture the friction of business collapses in real-time, often painting an overly optimistic picture of declining neighborhoods and contributing to the illusion of prosperity.

Recent studies have attempted to resolve this temporal latency by utilizing Street View Imagery (SVI) to capture the physical reality of streets \cite{Jiang2022, Zhang2019_SocialSensing, Xu2024, Ling2025}. Researchers have demonstrated that physical features—such as sky view factors and greenery—significantly correlate with vitality \cite{Long2016_Green}. Furthermore, the link between micro-scale walkability and human experience has been emphasized as a key driver for healthy urban living \cite{Liao2025, KOO2023}. More importantly, recent works have begun to use computer vision to identify attrition indicators, such as closed stores or roller shutters, providing a more realistic net vitality assessment. However, while the visual dimension has been significantly advanced, the semantic dimension of commercial quality remains fundamentally under-explored.

\subsection{From Visual Enumeration to Quality Diagnosis} 
Current SVI-based vitality assessments predominantly operate on a binary logic: distinguishing solely between open and closed storefronts. While detecting closures is a significant methodological stride, it simplifies all active stores into homogenous units. This binary approach ignores the substantial economic heterogeneity and varying resilience within active commerce.

For instance, a street lined with generic, subsistence-level retail shops may exhibit the same nominal ``openness ratio'' as a street dominated by high-end chain brands. Yet, their economic yield and risk-resistance capabilities are vastly different. Existing studies on storefront interfaces focus largely on physical transparency or signage density \cite{Mehta2009}, rarely penetrating the semantic content of the signage itself. Crucially, traditional Optical Character Recognition (OCR) methods fall short in this domain. They are often strictly literal, struggling with stylized logos and failing to resolve semantic ambiguities—such as linking multilingual aliases (e.g., mapping localized logograms to global brand names). The lack of semantic reasoning means that traditional models can quantify the mere presence of commerce but cannot evaluate its brand capital or market hierarchy.

To address these semantic bottlenecks, the field has seen a rapid adoption of Vision-Language Models (VLMs) for unstructured visual data processing. Recent systematic reviews confirm that VLMs have emerged as a robust tool for street view analytics and socioeconomic perception \cite{VLMReview2025}. State-of-the-art frameworks like UrbanVLP have successfully utilized multi-granularity VLM pretraining to predict broad urban indicators \cite{UrbanVLP2025}, while models such as MINGLE have extended VLM capabilities to detect semantically complex social regions \cite{MINGLE2025}. However, while these advanced models excel in general scene perception, they lack a dedicated economic mechanism for deciphering the precise brand hierarchy of commercial storefronts and coupling this intangible asset with continuous spatial externalities.

Our study bridges this gap by integrating VLMs with LLM-based semantic rectification. By doing so, we transition vitality assessment from basic visual enumeration to a robust qualitative diagnosis, establishing a systematic integration of hierarchical brand semantics and mall-induced spatial spillovers.

\subsection{From Rigid Buffers to Continuous Fields} 
In addition to internal storefront attributes, street vitality is profoundly shaped by exogenous anchors, particularly large-scale shopping malls. Existing literature acknowledges malls as vitality hubs but typically models their influence using simplistic spatial metrics, such as Euclidean distance to the nearest mall or rigid buffer zones \cite{Sevtsuk2014, Yue2017}.

These geometric approaches assume that commercial influence is uniform within a radius or decays linearly. They fail to capture the true agglomeration externalities of commercial centers—where the intensity of spatial spillover follows a non-linear distance decay law and varies significantly by the anchor's market position (e.g., a massive mixed-use complex radiates much further than a local supermarket). By ignoring these continuous spatial externalities, traditional metrics often underestimate or misjudge the economic potential of streets located in the spillover zones of major commercial centers. This study addresses this limitation by introducing a Gaussian decay function, modeling mall spillover as a continuous, typology-weighted energy field. This approach differentiates attenuation intensities based on fine-grained POI sub-categories, ensuring a realistic representation of commercial gravity and spatial resource distribution.

\section{Formulation}
Building on the visual-semantic framework, we formalize the Street Economic Vitality Index (SEVI) not merely as a statistical aggregation, but as a diagnostic feature matrix that captures the interplay between internal quality and external field effects. This section operationalizes the three dimensions—Commercial Activity, Spatial Utilization, and Physical Environment—into computable vectors, explicitly incorporating attrition filters (closure detection) and field-based interactions (spillover modeling) to provide the independent variables ($X$) for our subsequent spatiotemporal dynamic analysis.

\subsection{Mathematical Definition}
To ensure cross-indicator comparability, all raw variables are first directionally aligned (e.g., transforming the Closure Ratio into an openness proxy) and standardized using Min-Max normalization. For street segment $i$, we define the standardized indicator vector as:
\begin{equation}
\mathbf{x}_i = \begin{bmatrix} SD_{i} & 1 - CR_{i} & BR_{i} & MV_{i} & MD_{i} & ND_{i} & PP_{i} & GR_{i} & GD_{i} \end{bmatrix}^{\mathsf{T}} \in \mathbb{R}^{9} 
\end{equation}
Specifically, the components correspond to:
\begin{itemize}
    \item \textbf{Commercial Activity ($A_i$):} Shop Density ($SD_i$), Functional Occupancy Ratio ($1-CR_i$), Weighted Brand Ratio ($BR_i$), and Mall Spillover Vitality ($MV_i$). To minimize measurement error, the extraction of physical features—segmenting signboards, glass interfaces, and closed stores to calculate $SD_i$, $GD_i$, and $CR_i$ respectively—is rigorously executed using YOLOv5-seg prior to feeding the segmented signage into the VLM-LLM decoding pipeline for $BR_i$ calculation.
    \item \textbf{Spatial Utilization ($U_i$):} Motor Vehicle Density ($MD_i$), Non-motor Vehicle Density ($ND_i$), and Pedestrian Presence ($PP_i$).
    \item \textbf{Physical Environment ($P_i$):} Green Coverage Ratio ($GR_i$) and Storefront Glazing Density ($GD_i$).
\end{itemize}

These indicators are structured into three dimensions using a block-diagonal weight matrix, ensuring that indicators within each dimension are aggregated independently before final synthesis:
\begin{equation}
M =
\begin{bmatrix}
M_A & 0 & 0 \\
0 & M_U & 0 \\
0 & 0 & M_P
\end{bmatrix}
\in \mathbb{R}^{3\times 9}
\end{equation}
where $M_A \in \mathbb{R}^{1\times 4}$, $M_U \in \mathbb{R}^{1\times 3}$, and $M_P \in \mathbb{R}^{1\times 2}$ are the row vectors representing the weights calculated via the Entropy Weight Method (EWM) \cite{Zhu2020_EWM}, employed to avoid subjective bias and objectively capture the spatial information variance. Applying $M$ to $\mathbf{x}_i$ yields the three-dimensional feature vector $\mathbf{z}_i = [A_i, U_i, P_i]^{\mathsf{T}}$.

Finally, to generate a comprehensive diagnostic score for static spatial quality, the composite SEVI is derived via the TOPSIS method \cite{Hwang1981}. This ranks street segments by their relative proximity to the Ideal State ($\mathbf{z}^+$) and Negative State ($\mathbf{z}^-$):
\begin{equation}
\mathrm{SEVI}_i = \frac{d(\mathbf{z}_i, \mathbf{z}^-)}{d(\mathbf{z}_i, \mathbf{z}^+) + d(\mathbf{z}_i, \mathbf{z}^-)}
\end{equation}
This formulation ensures that the nine-dimensional diagnostic feature matrix accurately captures the built environment's vitality mechanisms, which are then individually regressed against real-time LBS crowd intensity to decode the temporal drivers of urban life. Meanwhile, the composite SEVI is utilized for macro-scale spatial diagnosis.

\subsection{Model Assumptions and Physical Interpretations}
The formulation rests on specific theoretical assumptions regarding urban mechanisms:
\begin{itemize}
    \item \textbf{Orthogonality of Dimensions:} While interactions exist, we treat Activity, Utilization, and Environment as analytically distinct layers to isolate specific urban deficits (e.g., identifying a street that is green but commercially dormant).
    
    \item \textbf{Category-Weighted Gaussian Field:} Unlike rigid buffer zones, we model Mall Spillover Vitality ($MV_i$) as a continuous energy field. Drawing on retail agglomeration theory \cite{Sevtsuk2014}, the influence of a commercial anchor $j$ decays non-linearly, mathematically constrained within a maximum spatial threshold $D = 2000m$:
    \begin{equation}
    MV_i = \sum_{j} \mathbb{I}(d_{ij} \leq D) \cdot \exp\left(-\frac{d_{ij}^2}{2\sigma_j^2}\right)
    \end{equation}
    Crucially, rather than relying on arbitrary parameter assignment, the attenuation bandwidth $\sigma_j$ is empirically calibrated using the Average Nearest Neighbor (ANN) distance for each commercial typology. Grounded in Central Place Theory, the spatial distribution density of a commercial facility inversely reflects its market reach and hierarchical level. For a mall $j$ belonging to category $c$ with $N_c$ total establishments in the city, $\sigma_j$ is defined as the mean Euclidean distance to its nearest competitor of the same tier:
    \begin{equation}
    \sigma_j = \frac{1}{N_c} \sum_{k=1}^{N_c} \min_{m \neq k} d(p_k, p_m)
    \end{equation}
    where $p_k$ and $p_m$ represent the spatial coordinates (projected to EPSG:3857 for metric accuracy) of malls within category $c$. Our empirical calibration reveals a rigorous spatial hierarchy: baseline retail facilities exhibit intense spatial competition with smaller energy fields (e.g., $\sigma = 991.84m$ for general markets and $\sigma = 1560.65m$ for standard shopping centers), whereas mixed-use complexes exhibit extensive spatial monopolies (e.g., $\sigma = 43045.21m$ for large-scale mall-residential hybrids). Consequently, given the 2000m spatial threshold, the mathematical interaction dictates that top-tier mega-malls generate a near-constant, saturated energy field within their immediate vicinity, whereas smaller retail facilities exhibit steep, continuous distance decay. This accurately mirrors the absolute spatial monopoly of large commercial anchors in the urban core. For singular establishments where ANN cannot be derived mathematically ($N_c < 2$), the baseline category mean is imputed to maintain distributional consistency. This data-driven calibration ensures that our model accurately reflects the true mechanics of commercial gravity and functional agglomeration premium. 
    
    However, we acknowledge a computational limitation in this spatial formulation. Relying on Euclidean distance for ANN calibration serves as a macro-scale heuristic, but it inherently simplifies the complex topological friction of urban environments. As highlighted in spatial interaction theory, physical barriers—such as rivers, arterial roads, or elevated highways—introduce significant urban friction that can sever pedestrian accessibility and distort actual spillover morphologies. While our strict 2,000-meter maximum threshold partially truncates this potential overestimation by confining the spillover within an immediate walking and local transit catchment, future iterations could integrate network-based shortest-path routing to capture these spatial discontinuities more rigorously.
    
    \item \textbf{Semantic Fidelity via VLM-LLM:} We posit that the semantic attributes extracted by VLMs—specifically brand hierarchy—serve as a high-fidelity proxy for economic quality. This assumes that the hierarchical weight of brands ($BR_i$) effectively filters the noise of low-value clustering and reflects the actual economic resilience of the street interface better than raw counting metrics.
\end{itemize}

\section{Materials and Methods}
We established a comprehensive analytical framework to diagnose the economic vitality of urban streets. Moving beyond traditional single-source metrics, this study constructs a multimodal data fusion pipeline that integrates high-resolution Street View Imagery (SVI) with semantic geospatial data. This section details the study area, the visual-semantic interpretation pipeline, and the operationalization of vitality indicators.

\subsection{Study Area} 
\begin{figure}[htbp]
    \centering
    \includegraphics[width=0.5\textwidth]{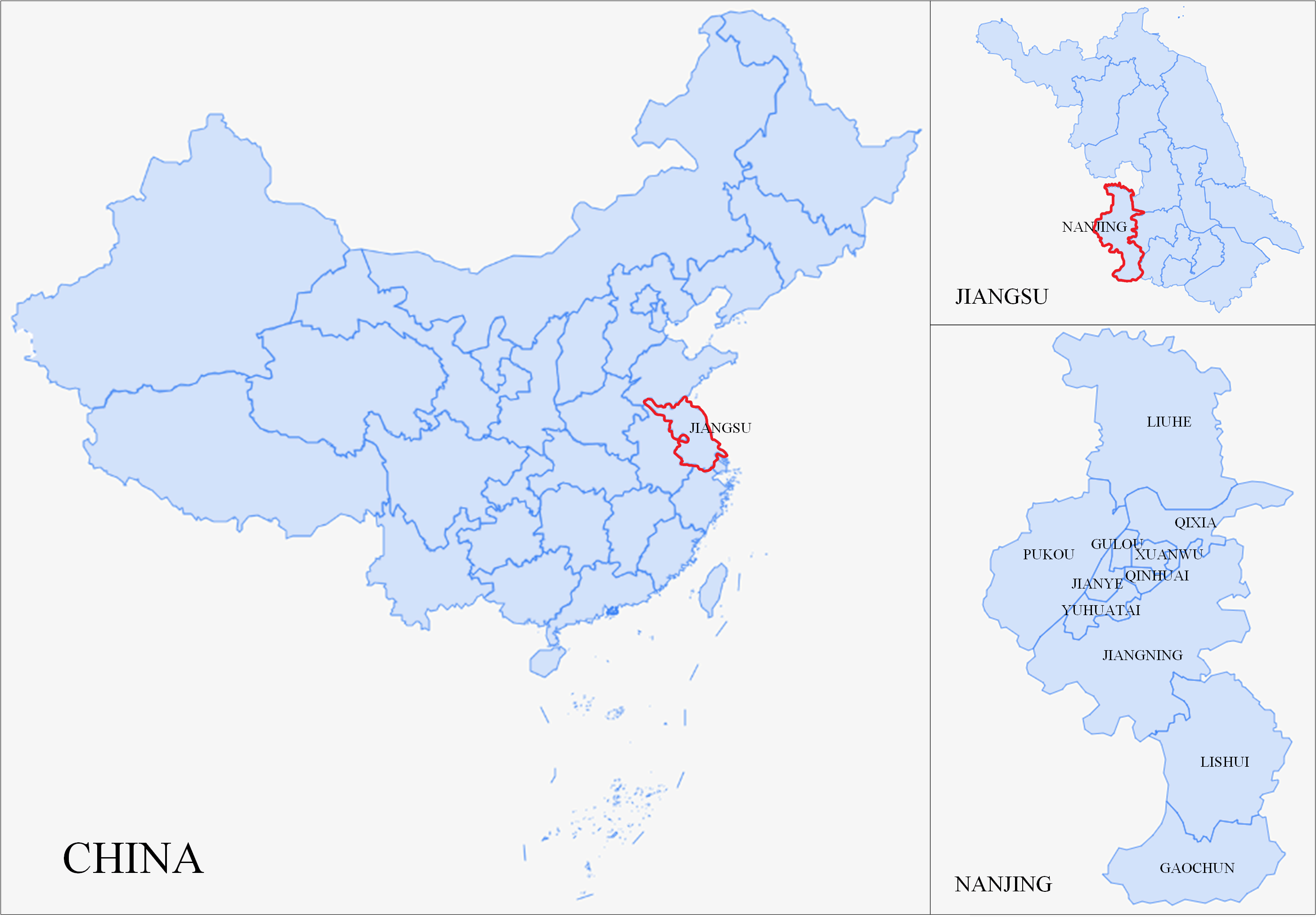}
    \caption{Study area.}
    \Description{A map showing the central metropolitan area of Nanjing, Jiangsu Province, which serves as the study area.}
    \label{FIG:1}
\end{figure}

The study focuses on the central metropolitan area of Nanjing, Jiangsu Province (Figure~\ref{FIG:1}). As a city with over 2,500 years of history, Nanjing exemplifies the typical challenges of modern urban renewal: a complex mosaic of historical preservation zones, aging residential neighborhoods, and high-density commercial cores. While the master plan identifies this area as the primary axis for vitality, it faces significant spatial fragmentation \cite{NanjingPlan2024}. The coexistence of thriving commercial hubs and declining, hollowing-out streets makes it an ideal testbed for our micro-scale diagnostic framework. Validating SEVI in such a heterogeneous environment ensures its transferability to other dense urban contexts facing similar regeneration pressures.

\subsection{Data Acquisition and Processing Framework}
\begin{figure}[htbp]
    \centering
    \includegraphics[width=1.0\textwidth]{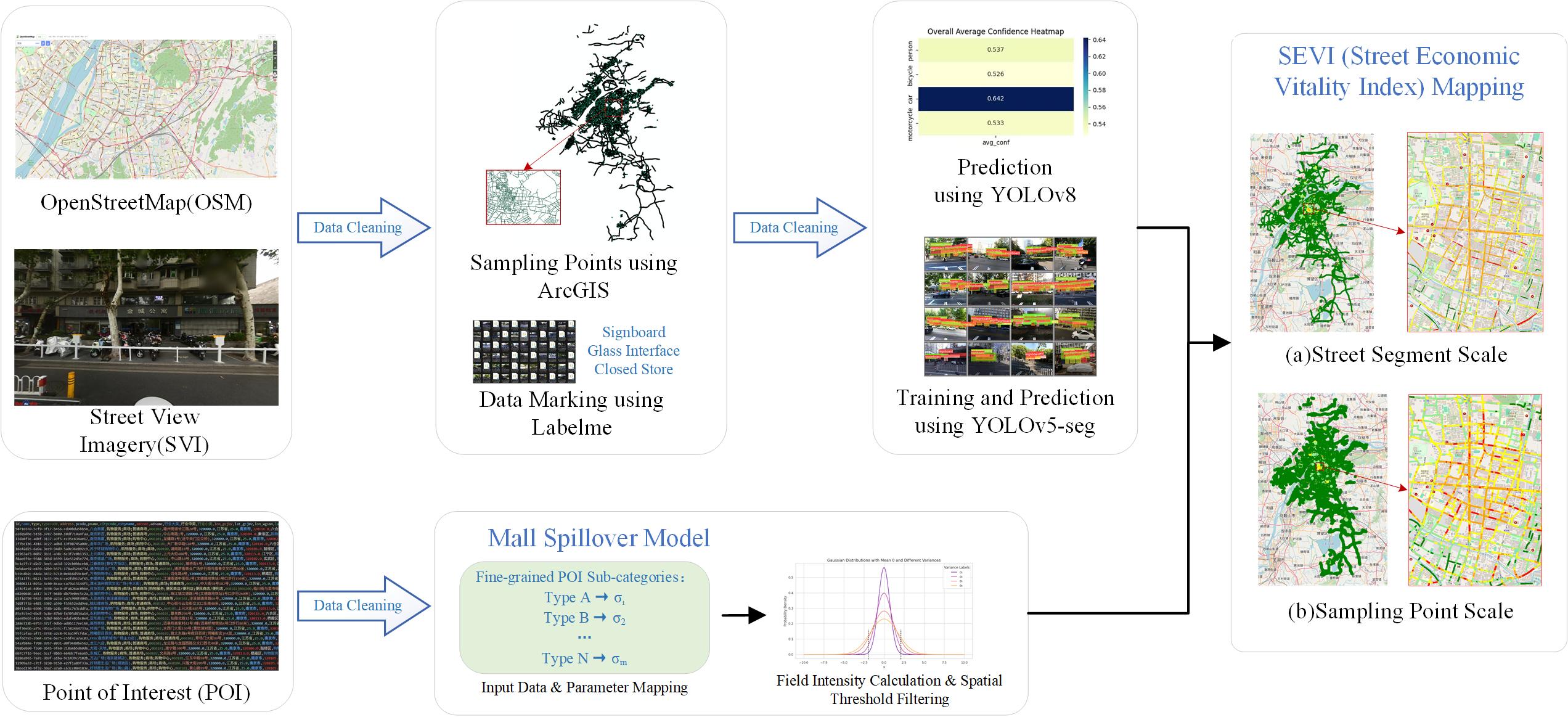}
    \caption{Analysis framework.}
    \Description{A flowchart illustrating the multimodal data fusion pipeline, from SVI acquisition to visual-semantic interpretation.}
    \label{FIG:2}
\end{figure}

To capture the fine-grained texture of street life, we implemented a systematic sampling and acquisition workflow (Figure~\ref{FIG:2}). We first constructed the foundational road network by extracting pedestrian-accessible segments from OpenStreetMap, excluding highways and tunnels to yield 7,153 valid segments. To ensure a continuous visual narrative, we adopted a high-frequency sampling strategy at 20-meter intervals—a significant resolution upgrade compared to the sparse 50m or 100m sampling intervals commonly used in previous studies \cite{Li2022, Jiang2022}. This density is critical for capturing micro-variations in storefront continuity and detecting localized dead spots that coarser sampling might miss. Consequently, using the Baidu Maps API, we retrieved 557,672 panoramic images (1024 $\times$ 800 pixels) across 278,836 unique points, with each point capturing dual-directional views ($90^\circ$ and $270^\circ$ relative to the road heading) to cover the street interface comprehensively.

Crucially, to avoid simultaneity bias and explicitly introduce causal awareness, we implemented a strict temporal lag design between the built environment features and vitality outcomes. The explanatory variables ($X_{i, t-1}$) are derived from a synthesized historical baseline comprising the SVI data (captured in or before 2022) and finely curated Point of Interest (POI) data from 2023. This integration is methodologically intentional: SVI effectively records "slow variables" such as building facades, structural closures, and green infrastructure, while the 2023 POI data provides a necessary semantic update to capture the "fast variables" of commercial turnover. Together, they constitute the accumulated physical and commercial baseline of the street at time $t-1$.

\begin{figure}[htbp]
    \centering
    \includegraphics[width=1.0\textwidth]{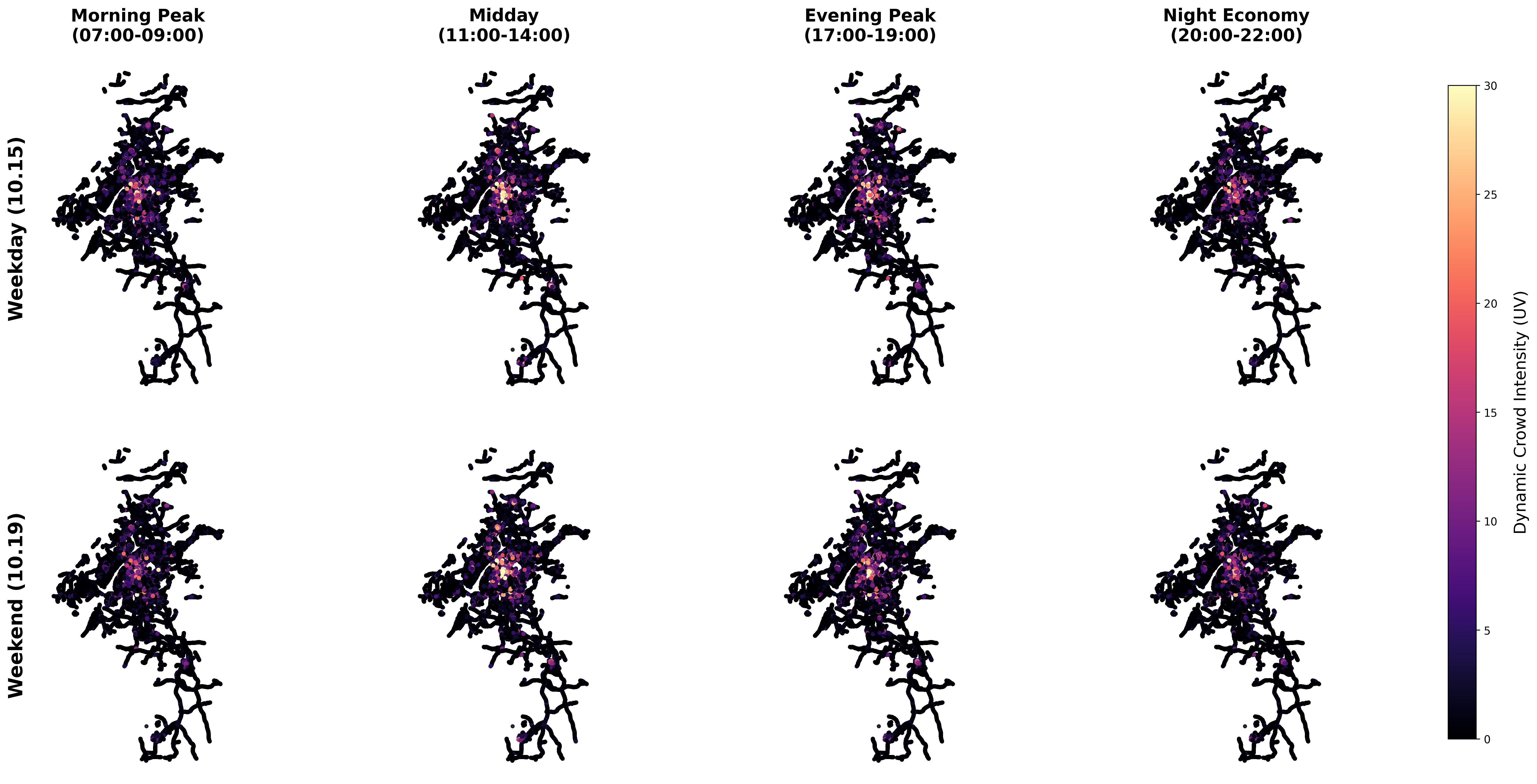}
    \caption{Spatiotemporal evolution of dynamic crowd intensity (UV) across four tidal periods for both a weekday and a weekend in Nanjing.}
    \Description{A 2x4 matrix of heatmaps showing the dynamic crowd intensity across different tidal periods. The hotspots shift and expand significantly during the midday and night economy periods compared to the morning peak.}
    \label{FIG:3}
\end{figure}

Furthermore, to overcome the snapshot limitation inherent in static visual audits, this study incorporates dynamic Location-Based Services (LBS) data from October 2025 as the primary proxy for actual street vitality (the dependent variable, $Y_{i, t}$). This natural temporal lag ($t-1$ preceding $t$) logically isolates the pre-existing spatial supply from subsequent human behavior, ensuring that our model captures the quasi-causal driving effects of historical street morphologies rather than purely descriptive co-occurrences. Sourced from mobile signaling and location heatmaps, the LBS dataset captures real-time active user volumes (UV), representing the revealed preference and realized spatial demand of urban residents. To map the spatiotemporal rhythms of urban life, the raw data was temporally aggregated into four distinct tidal periods: Morning Peak (07:00--09:00), Midday (11:00--14:00), Evening Peak (17:00--19:00), and Night Economy (20:00--22:00), covering both a typical weekday (October 15) and a weekend (October 19). Spatially, the raw positioning data was cleaned to remove signal drift noise and spatially joined to the 7,153 street segments. As illustrated in Figure~\ref{FIG:3}, the dynamic crowd intensity exhibits profound spatiotemporal heterogeneity, revealing a distinct tidal pattern that necessitates a time-sliced modeling approach to decouple the underlying spatial drivers across different periods.

\subsection{The Visual-Semantic Analysis Pipeline} 
A core innovation of this study is the shift from visual detection to semantic understanding. We developed a two-stage deep learning pipeline to extract both the physical status and semantic quality of street entities.

\subsubsection{Diagnostic Object Segmentation}
To quantify indicators, we defined a taxonomy of visual markers specifically relevant to urban renewal (Table~\ref{tbl1}) and employed YOLOv5-seg \cite{Redmon2016_YOLO} for instance segmentation. The model was trained to identify and segment three distinct classes: signboards, glass interfaces, and closed stores. Unlike simple object detection, instance segmentation generates pixel-level masks, enabling the precise separation of these elements from complex backgrounds. These segmented markers serve specific downstream analytical functions: signboards are quantified to measure Shop Density, while glass interfaces are used to calculate Shopfront Glazing Density. Crucially, the joint detection of closed stores and signboards allows for the computation of the Closure Ratio, thereby capturing the negative commercial dynamics often missed by traditional POI data. Furthermore, the extracted signboard regions serve as the direct input for our subsequent VLM-LLM Brand Decoding Pipeline.

\begin{figure}[htbp]
    \centering
    \includegraphics[width=1.0\textwidth]{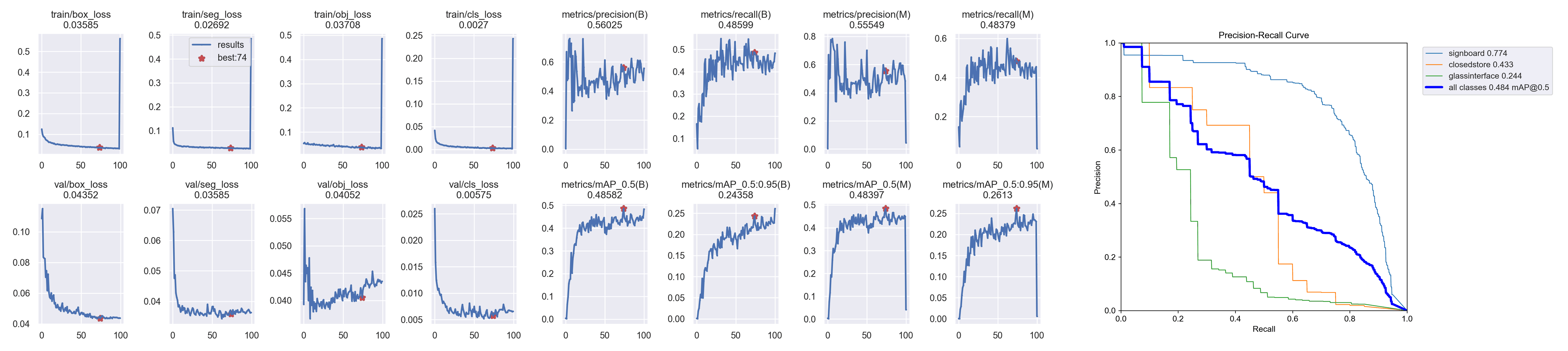}
    \caption{Training performance of YOLOv5-seg.}
    \Description{Line charts displaying the loss and mAP training curves over 100 epochs for the YOLOv5-seg model.}
    \label{FIG:4}
\end{figure}

The model was trained on a manually annotated dataset of 642 images using the LabelMe tool \cite{Wada2021} (parameters in Table~\ref{tbl2}). As illustrated in the training performance curves (Figure~\ref{FIG:4}), the model exhibited stable convergence characteristics. Ultimately, the detection of Signboards achieved high precision ($mAP@0.5 = 0.774$), providing a reliable foundation for the subsequent semantic analysis.

\subsubsection{Semantic Brand Decoding via VLM-LLM Pipeline}
\begin{figure}[htbp]
    \centering
    \includegraphics[width=1.0\textwidth]{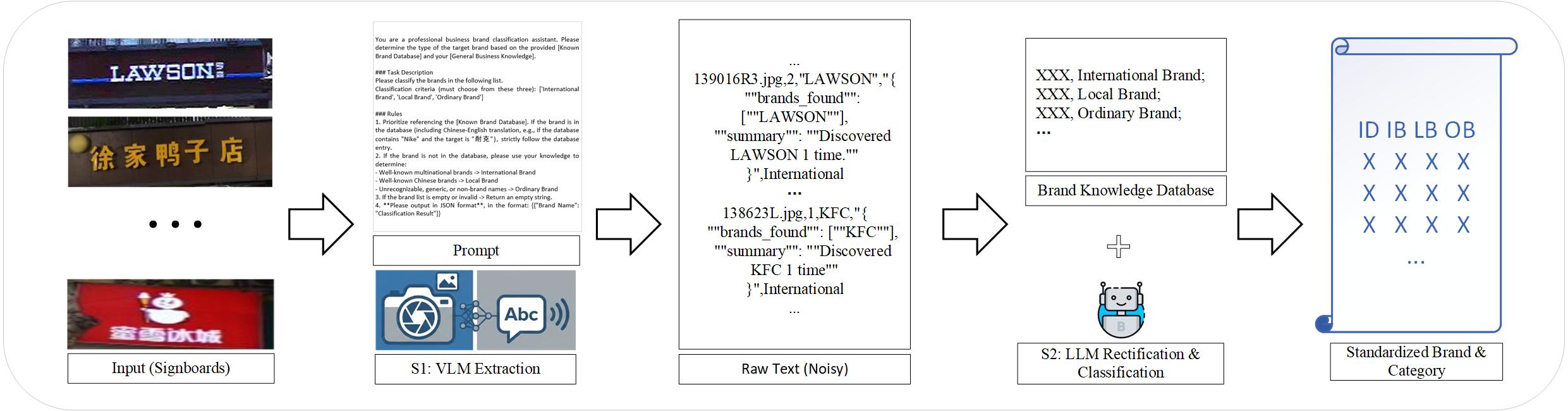}
    \caption{The architecture of the VLM-LLM Brand Decoding Pipeline.}
    \Description{A diagram showing the two-stage VLM-LLM pipeline for extracting and refining brand semantics from storefront images.}
    \label{FIG:5}
\end{figure}

To address the limitation that all active stores look alike in traditional metrics, we developed a Visual-Language Model (VLM) driven Brand Decoding module. As illustrated in Figure~\ref{FIG:5}, the pipeline leverages a dual-stage recognition-rectification framework to process street-view imagery:
   
(1) \textbf{Stage 1: VLM Extraction (S1).} We employ a pre-trained VLM (Qwen2-VL-7B-Instruct) to transcribe textual information and identify commercial logos from the input images. To minimize model hallucination and ensure deterministic outputs, generation hyperparameters were strictly constrained (\texttt{temperature = 0.0}, \texttt{top\_p = 1.0}, \texttt{max\_new\_tokens = 64.0}, and \texttt{do\_sample = False}). The VLM is guided by a carefully engineered zero-shot prompt instructing it to act as a street-view signage recognition assistant. To explicitly suppress hallucination, the prompt enforces strict constraints (i.e., extracting only visible text/logos without fabrication). The output is mandated as a structured JSON object containing \texttt{brands\_found} and a descriptive \texttt{summary}. If no clear brands are present, it returns an empty list, effectively filtering out environmental noise.

(2) \textbf{Stage 2: LLM Rectification \& Classification (S2).} The raw, potentially noisy text list extracted from S1 is fed into an LLM agent powered by DeepSeek-Chat (\texttt{temperature = 0.0}, \texttt{top\_p = 1.0}, \texttt{max\_new\_tokens = 128.0}, \texttt{do\_sample = False}). To standardize heterogeneous inputs, the LLM acts as a professional commercial classification assistant, bridging local observations with global semantic hierarchies. The prompt dynamically injects both a customized reference database and the raw VLM outputs, enforcing a strict reasoning hierarchy to classify brands into International Brand, Local Brand, or Ordinary Brand tiers. The final output is strictly constrained to a key-value JSON format, ensuring seamless programmatic integration for the subsequent calculation of the Weighted Brand Ratio ($BR_i$). Finally, to mitigate spatial discontinuity inherent in discrete sampling points, we apply a spatial sliding window (size=5) to smooth the $BR_i$ index, ensuring it reflects the continuous commercial atmosphere.

\subsubsection{Validation and Ablation of the Semantic Pipeline}
To validate the efficacy of the proposed dual-stage visual-semantic pipeline and quantify the necessity of LLM rectification, we conducted an ablation study. We constructed a manually annotated ground truth (GT) dataset comprising 200 street view images, categorized into 99 International Brands, 54 Ordinary Brands, and 47 Local Brands. We evaluated three progressive pipeline configurations: Traditional OCR (EasyOCR + RapidFuzz), VLM Only (Qwen2-VL), and VLM+LLM (Ours).

As presented in Table \ref{tab:ablation}, the traditional OCR approach struggles significantly with the stylized fonts, occlusions, and complex lighting inherent in streetscapes, yielding an overall $F1$-score of only 0.412. Upgrading the visual backbone to a VLM markedly improves text recognition ($F1 = 0.652$). Crucially, the introduction of the LLM rectification agent standardizes localized semantic variants and corrects extraction noise, elevating the overall $F1$-score to 0.821. Notably, the VLM+LLM pipeline achieves an $F1$-score of 0.920 for Ordinary Brands and 0.768 for International Brands, representing a 99.2\% relative improvement in overall accuracy compared to the OCR baseline. This quantitative leap justifies our claim that high-fidelity brand hierarchy extraction requires deep semantic reasoning beyond primitive OCR, laying a highly accurate data foundation for subsequent spatial vitality modeling.

\begin{table}[htbp]
\centering
\caption{Performance comparison of brand recognition pipelines (Ablation Study)}
\label{tab:ablation}
\resizebox{\textwidth}{!}{
\begin{tabular}{lccccc}
\toprule
\textbf{Brand Type} & \textbf{GT Count} & \textbf{Metric} & \textbf{OCR} & \textbf{VLM Only} & \textbf{VLM + LLM (Ours)} \\
\midrule
International & 99 & Precision & 0.474 & 0.638 & \textbf{0.802} \\
 &  & Recall & 0.374 & 0.606 & \textbf{0.737} \\
 &  & $F1$-score & 0.418 & 0.622 & \textbf{0.768} \\
\midrule
Ordinary & 54 & Precision & 0.449 & 0.760 & \textbf{1.000} \\
 &  & Recall & 0.407 & 0.704 & \textbf{0.852} \\
 &  & $F1$-score & 0.427 & 0.731 & \textbf{0.920} \\
\midrule
Local & 47 & Precision & 0.425 & 0.694 & \textbf{0.939} \\
 &  & Recall & 0.362 & 0.532 & \textbf{0.660} \\
 &  & $F1$-score & 0.391 & 0.602 & \textbf{0.775} \\
\midrule
\textbf{Overall (Mean)} &   & Precision & 0.449 & 0.698 & \textbf{0.914} \\
 &  & Recall & 0.381 & 0.614 & \textbf{0.750} \\
 &  & $F1$-score & 0.412 & 0.652 & \textbf{0.821} \\
\bottomrule
\end{tabular}}
\end{table}

\subsubsection{Mobility and Environment Sensing}
To complement the storefront analysis, we employed a pre-trained YOLOv8-m model \cite{Jocher2023} to detect dynamic mobility flows. The widely-used COCO dataset classes \cite{Lin2014} were mapped to our mobility indicators. The model achieved robust detection performance ($mAP@0.5 \approx 0.64$ for mobility classes), ensuring reliable estimation of street-level utilization intensity.

\subsection{Indicator Calculation and Operationalization}
Based on the visual and semantic data extracted above, we operationalized the nine indicators defined in the Formulation section. Table~\ref{tbl3} summarizes the calculation logic. The operationalization of Mall Spillover Vitality ($MV_i$) directly follows the empirical Average Nearest Neighbor (ANN) calibration defined in Section 2, ensuring that the continuous spatial externalities are rigorously constrained within the 2,000-meter threshold. Ultimately, these nine carefully engineered static spatial features will be regressed against the dynamic LBS crowd intensity ($Y$) to decode the temporal heterogeneity of street vitality.

\begin{table}[htbp]
\centering
\caption{Annotated visual categories and their renewal-oriented interpretations.} 
\label{tbl1}
\begin{tabular}{ p{1.8cm} p{2.5cm} p{6.5cm} }
\toprule
\textbf{Category} & \textbf{Visual Example} & \textbf{Description} \\
\midrule
Signboard & 
\vspace{0pt}\includegraphics[width=2.2cm]{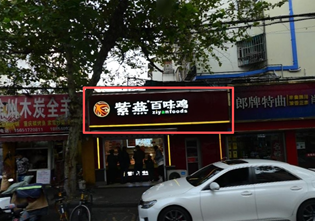} & 
A proxy for commercial density and the primary carrier of semantic brand information, reflecting the qualitative hierarchy of local businesses. \\
\midrule
Closed Store & 
\vspace{0pt}\includegraphics[width=2.2cm]{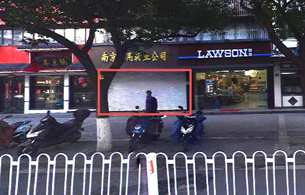} & 
A visual indicator of economic attrition; captures vacancy dynamics and hidden recession often missed by static POI datasets. \\
\midrule
Glass Interface & 
\vspace{0pt}\includegraphics[width=2.2cm]{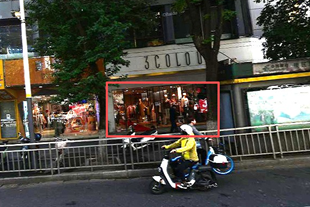} & 
Represents the permeability of the street canyon, facilitating visual interaction between indoor commerce and outdoor pedestrians. \\
\midrule
Person & 
\vspace{0pt}\includegraphics[width=2.2cm]{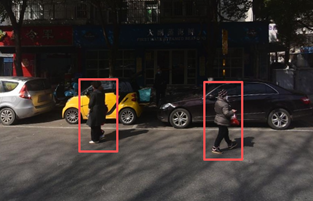} & 
Direct proxy for social vibrancy; reflects the street's attractiveness and capacity for social interaction. \\
\midrule
Bicycle & 
\vspace{0pt}\includegraphics[width=2.2cm]{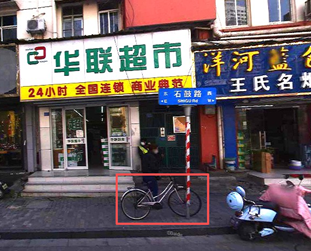} & 
Reflects sustainable mobility patterns and green accessibility. \\
\midrule
Car & 
\vspace{0pt}\includegraphics[width=2.2cm]{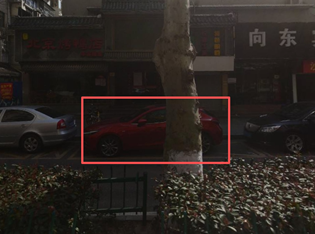} & 
Indicates vehicular accessibility and commercial logistics intensity. \\
\midrule
Motorcycle & 
\vspace{0pt}\includegraphics[width=2.2cm]{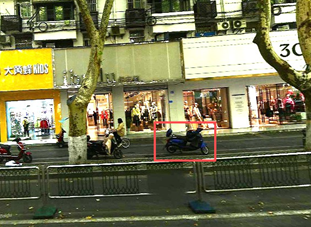} & 
Reveals informal transport dynamics and delivery activities that shape street vitality. \\
\bottomrule
\end{tabular}
\end{table}

\begin{table}[htbp]
\centering
\caption{Training Parameters}\label{tbl2}
\begin{tabular}{ll}
\toprule
\textbf{Index} & \textbf{Value} \\
\midrule
Hardware Environment & 5 $\times$ NVIDIA GeForce RTX 3090 (Data Parallel) \\
Batch size     & 8 \\
Epochs         & 100 \\
Training images & 642 
\\
Learning rate  & 0.01 \\
\bottomrule
\end{tabular}
\end{table}

\begin{table}[htbp]
\renewcommand{\arraystretch}{1.5} 
\scriptsize
\centering
\caption{Summary of variables, calculation formulas, and definitions.}
\label{tbl3}
\begin{tabular}{p{2.2cm} p{4.2cm} p{4.8cm}}
\toprule
\textbf{Variable} & \textbf{Formula} & \textbf{Definition of Components} \\
\midrule
Shop Density & 
$SD_{i} = \frac{NS_{L_i} + NS_{R_i}}{l_i}$ & 
$NS$: Number of signboards detected on the respective side. \\

Closure Ratio & 
$CR_{i} = \frac{NC_{L_i} + NC_{R_i}}{NS_{L_i}+NS_{R_i}}$ & 
$NC$: Number of closed stores (identified by roller shutters/locked gates). \newline
$NS$: Total count of signboards (active + closed). \\

Weighted Brand Ratio & 
$BR_{i} = \frac{NB^w_{L_i} + NB^w_{R_i}}{NS_{L_i} + NS_{R_i}}$ & 
$NB^w$: Weighted brand score (numerator). \newline
Calculation: $NB^w = N_{loc} \cdot w_{1} + N_{glb} \cdot w_{2}$ \newline
$N_{loc/glb}$: Number of Local/International brands. \newline
$w_{1}, w_{2}$: Hierarchy weights (e.g., $1.0$ vs $1.5$). \newline
\textit{Note:} Using the total signboard count (NS) as the denominator inherently penalizes streets with high closure rates, reflecting the dilution of brand premium in decaying corridors. \\

Mall Spillover Vitality & 
$MV_i = \sum_{j=1}^{N} \mathbb{I}(d_{ij} \leq D) \cdot \exp\left(-\frac{d_{ij}^2}{2\sigma_j^2}\right)$ & 
$d_{ij}$: Euclidean distance between street unit $i$ and mall $j$. \newline
$\sigma_j$: Gaussian bandwidth specific to mall $j$'s hierarchy. \newline
$\mathbb{I}(\cdot)$: Indicator function (1 if $d_{ij} \leq D$, 0 otherwise). \newline
$D$: Max spatial threshold (e.g., 2000m). \\

\midrule
Motor Vehicle Density & 
$MD_{i} = \frac{NM_{L_i} + NM_{R_i}}{l_i}$ & 
$NM$: Number of detected motor vehicles (cars, trucks, buses). \\

Non-Motor Vehicle Density & 
$ND_{i} = \frac{NN_{L_i} + NN_{R_i}}{l_i}$ & 
$NN$: Number of non-motor vehicles (bicycles, e-scooters). \\

Pedestrian Presence & 
$PP_{i} = \frac{NP_{L_i} + NP_{R_i}}{l_i}$ & 
$NP$: Number of pedestrians observed on the street. \\

\midrule
Green Coverage Ratio & 
$GR_{i} = \frac{NG^{pix}_{L_i} + NG^{pix}_{R_i}}{NT^{pix}_{L_i} + NT^{pix}_{R_i}}$ & 
$NG^{pix}$: Number of green vegetation pixels. \newline
$NT^{pix}$: Total pixel count of the panoramic image area. \\

Shopfront Glazing Density & 
$GD_{i} = \frac{NG^{inst}_{L_i} + NG^{inst}_{R_i}}{{l_i}}$ & 
$NG^{inst}$: Number of detected glass interface instances. \\

\bottomrule
\end{tabular}
\vspace{1.5ex}
{\tiny \textit{Note:} For all equations, subscripts $L_i$ and $R_i$ denote the Left and Right sides of street segment $i$, respectively. $l_i$ denotes the length of street segment $i$ (in meters). All counts are derived from dual-directional SVI sampling points aggregated to the segment level.\par}
\end{table}

\section{Variables Description}

To systematically diagnose the multifaceted nature of street-level economics and explain the spatiotemporal heterogeneity of the dynamic LBS crowd intensity (the dependent variable), we operationalized nine independent spatial features into a tri-dimensional diagnostic matrix (Table~\ref{tbl3}). The statistical distributions of these static indicators at the sampling-point level are visualized in Figure~\ref{FIG:6}, revealing significant micro-scale spatial heterogeneity across the study area. Unlike traditional metrics that focus solely on volume, our indicator system is designed to capture the nuance among quantity (density), quality (brand and environment), and status (market exit vs. active).

\subsection{Commercial Activity}
\begin{figure}[htbp]
    \centering
    \includegraphics[width=0.9\textwidth]{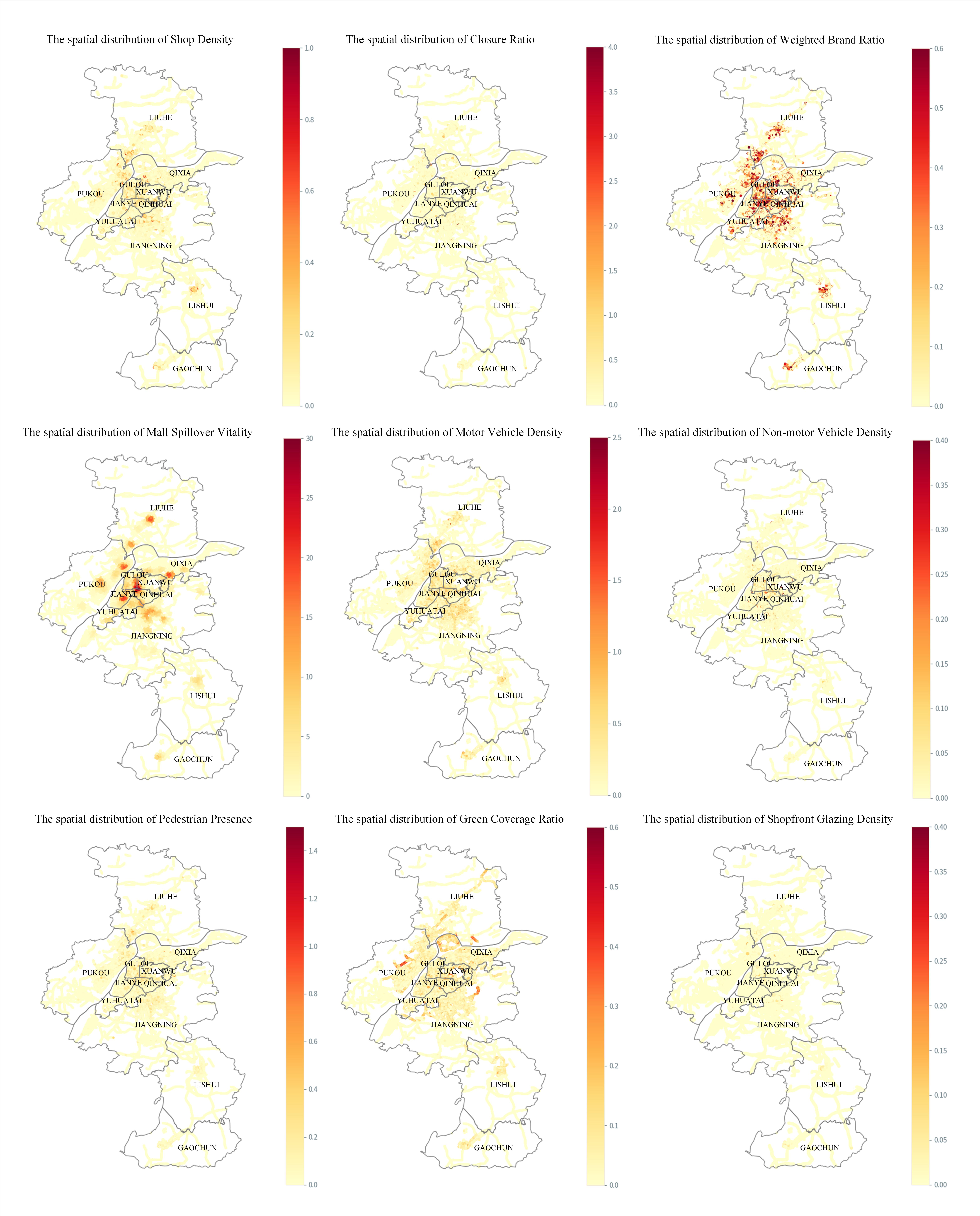}
    \caption{The distribution of variables at the sampling-point level.}
    \Description{A series of maps showing the spatial distribution of all nine vitality variables across the sampling points.}
    \label{FIG:6}
\end{figure}

\begin{figure}[htbp]
    \centering
    \includegraphics[width=0.4\textwidth]{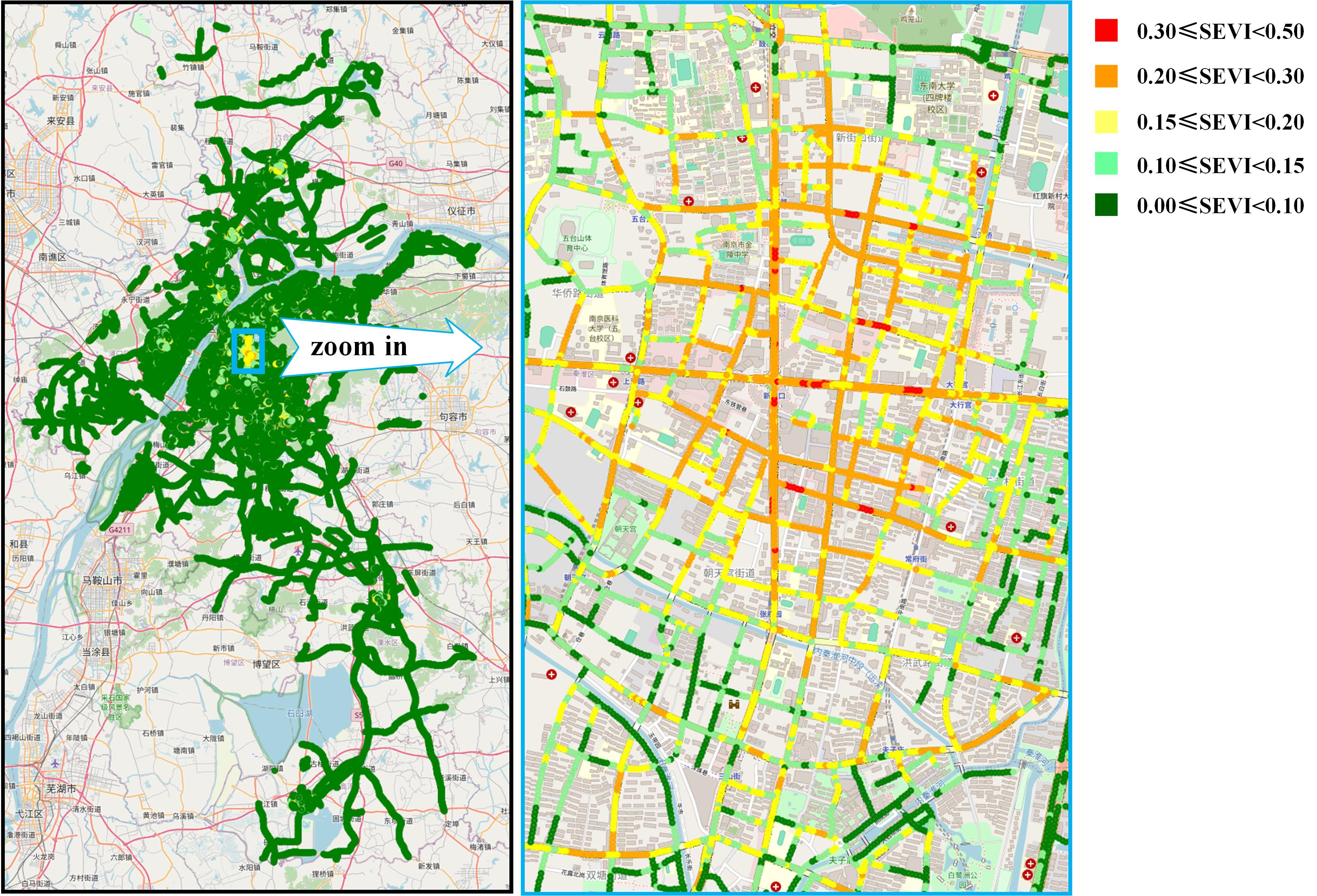}
    \caption{The distribution of Commercial Activity at the sampling-point level.}
    \Description{A spatial distribution map illustrating the concentration of Commercial Activity indicators at the sampling-point level.}
    \label{FIG:7}
\end{figure}

This dimension serves as the core supply-side driver of street vitality; however, we posit that mere establishment density is insufficient to reflect true economic prosperity. To construct a robust assessment, we synthesize four complementary indicators (individual spatial distributions detailed in Figure~\ref{FIG:6}). First, Shop Density ($SD_{i}$) measures the absolute concentration of commercial entities per unit length, providing a baseline assessment of commercial agglomeration. Second, acting as a critical attrition filter, the Closure Ratio ($CR_{i}$) visually identifies closed storefronts. This variable explicitly corrects the survivorship bias inherent in traditional POI data, distinguishing between streets that are functionally active and those suffering from economic distress; thus, $1-CR_{i}$ represents the effective functional survival rate. Third, the Weighted Brand Ratio ($BR_{i}$) transcends simple binary enumeration to capture the intangible brand premium of commerce. Derived from the VLM-LLM semantic pipeline, this metric assigns differential weights to international chains and local brands, effectively distinguishing modernized commercial hubs from generic, subsistence-level retail clustering. Finally, Mall Spillover Vitality ($MV_{i}$) quantifies the continuous spatial externalities of large commercial anchors. As detailed in the methodology, this category-weighted Gaussian field captures how large complexes export commercial gravity to surrounding streets. The aggregated spatial distribution of this dimension (Figure~\ref{FIG:7}) clearly highlights the structural dependency of local street vitality on regional hubs, forming continuous high-activity corridors in the urban core.

\subsection{Spatial Utilization}
\begin{figure}[htbp]
    \centering
    \includegraphics[width=0.4\textwidth]{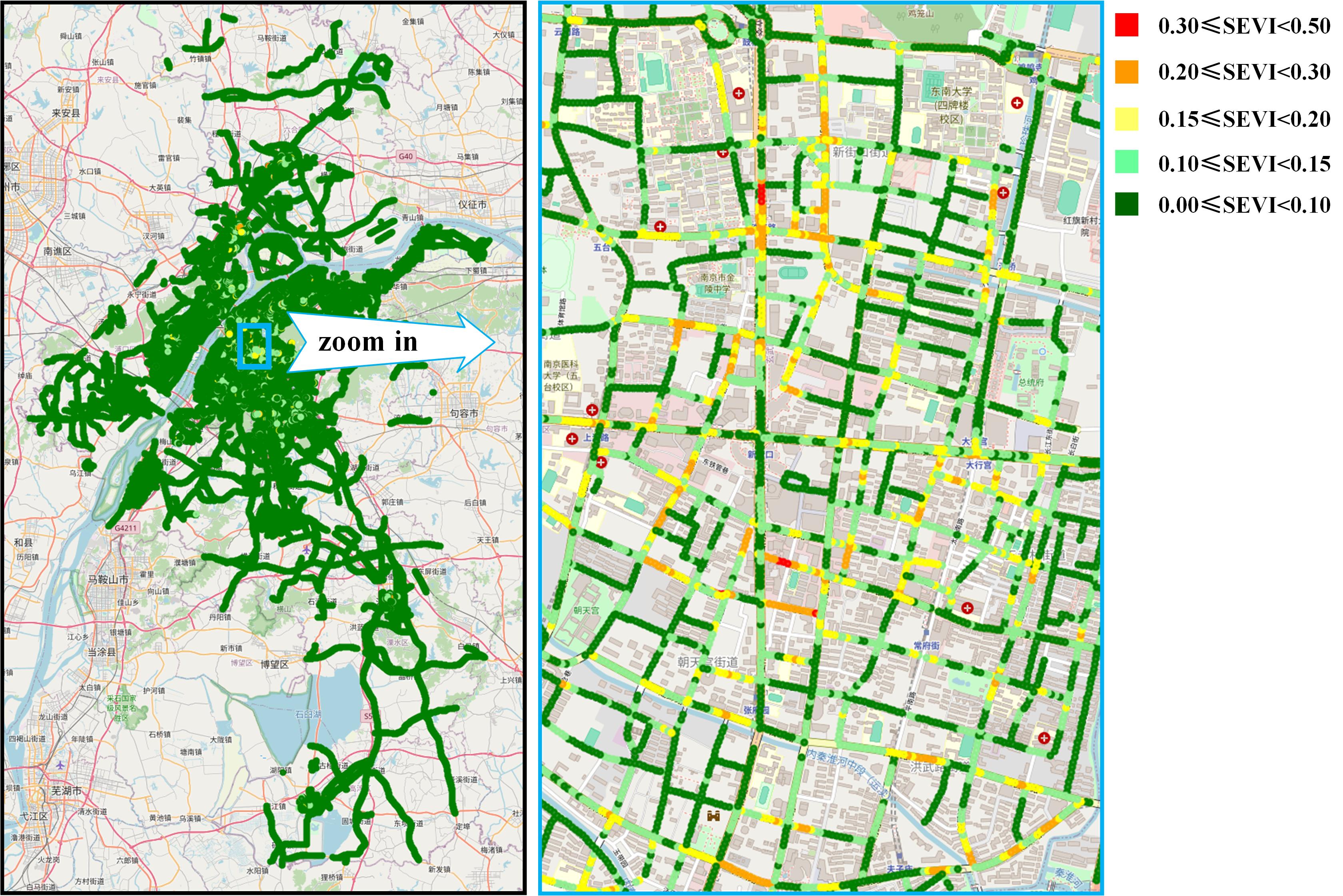}
    \caption{The distribution of Spatial Utilization at the sampling-point level.}
    \Description{A spatial distribution map illustrating the concentration of Spatial Utilization indicators at the sampling-point level.}
    \label{FIG:8}
\end{figure}

Urban vitality is sustained by the continuous flow of economic agents. We measure the dynamic pulse of the street through three mobility indicators. Pedestrian Presence ($PP_{i}$) reflects the intensity of human-scale engagement and serves as the most direct proxy for revealed consumer demand, echoing classic theories of life between buildings \cite{Gehl1987}. Complementing this, Motor Vehicle Density ($MD_{i}$) and Non-motor Vehicle Density ($ND_{i}$) capture spatial accessibility and transit-oriented activity. These indicators reveal the critical trade-offs between motorized transit efficiency and pedestrian-friendly vibrancy—a spatial friction often observed in the design of complete streets. Mapping these flows (Figure~\ref{FIG:8}) aids in identifying spatial bottlenecks where commercial supply mismatches mobility demand.

\subsection{Physical Environment}
\begin{figure}[htbp]
    \centering
    \includegraphics[width=0.4\textwidth]{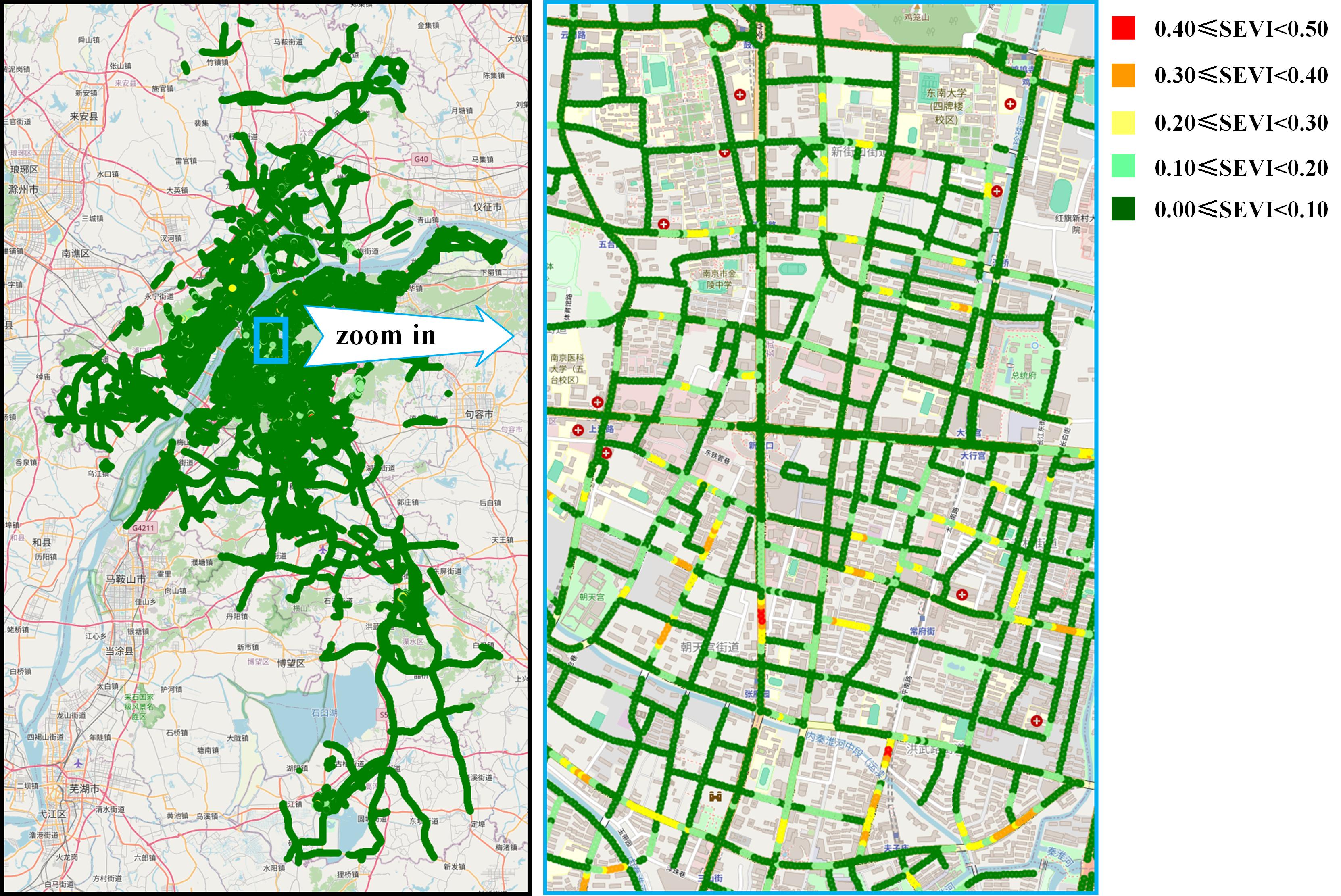}
    \caption{The distribution of Physical Environment at the sampling-point level.}
    \Description{A spatial distribution map illustrating the concentration of Physical Environment indicators at the sampling-point level.}
    \label{FIG:9}
\end{figure}

The built environment determines the comfort, amenity value, and interaction potential of the street canyon. We focus on two key interface qualities. Shopfront Glazing Density ($GD_{i}$) measures the visual permeability of the interface. High transparency facilitates visual interaction between indoor commercial supply and outdoor pedestrian demand, effectively reducing information asymmetry—a key design principle aligned with the hedonic value of walkability \cite{Jacobs1961, Mehta2009}. Meanwhile, the Green Coverage Ratio ($GR_{i}$) reflects the provision of ecological amenities and visual comfort. While occasionally exhibiting a negative spatial correlation with extreme commercial density, environmental amenity remains an indispensable factor for long-term urban resilience and livability \cite{Long2016_Green, Ye2019_LUP}. Visualizing this dimension (Figure~\ref{FIG:9}) highlights the structural disparity between hard commercial streets and soft livable neighborhoods.

\subsection{Dual-Scale Diagnostic Capabilities} 
A distinguishing feature of this framework is its scalar flexibility, bridging macro-level resource allocation with micro-level spatial intervention \cite{Huang2025}. At the street-segment level (macro-diagnosis), aggregating scores provides a macroscopic view suitable for district-level master planning and identifying regional corridors of economic decline. Conversely, at the sampling-point level (micro-surgery), preserving data at 20-meter intervals captures fine-grained spatial heterogeneity. This ultra-high resolution is critical for precision urban renewal, enabling policymakers to pinpoint localized market failures (e.g., a 100-meter stretch of closed shops or low-quality facades) within an otherwise healthy economic corridor, thereby guiding targeted spatial investments.

\section{Results}

\subsection{Unveiling the Logic of Vitality}
\begin{figure}[htbp]
    \centering
    \includegraphics[width=0.6\textwidth]{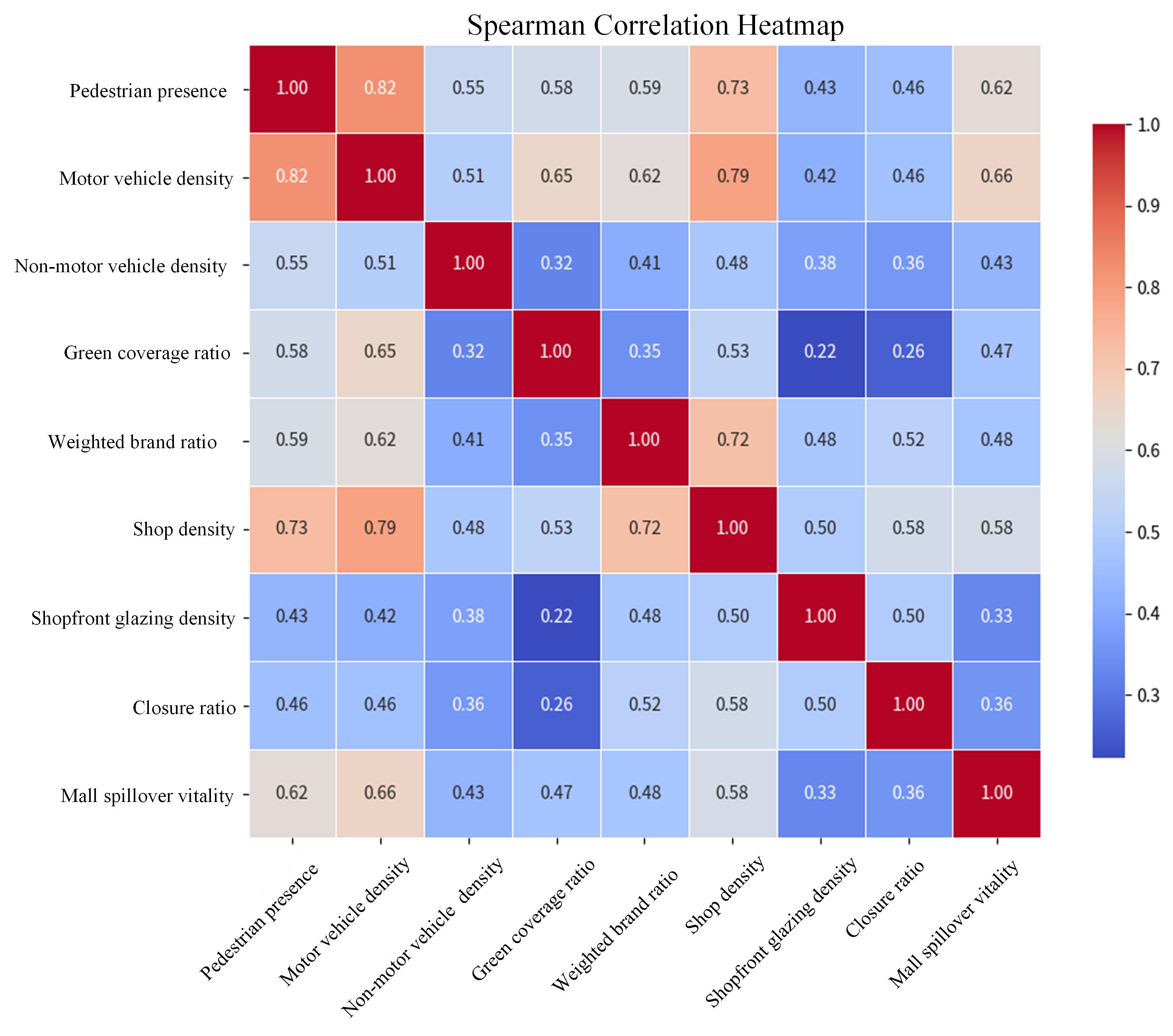}
    \caption{Spearman correlation heatmap identifying the internal coupling of vitality indicators.}
    \Description{A Spearman correlation heatmap showing the relationships among the nine vitality indicators.}
    \label{FIG:10}
\end{figure}

Spearman rank correlation analysis unveils the internal coupling mechanisms among street vitality components, as illustrated in Figure~\ref{FIG:10}.

The analysis first highlights a robust Traffic-Commerce Nexus, where Motor Vehicle Density ($MD_i$) and Shop Density ($SD_i$) exhibit a strong positive correlation ($r=0.79$). This empirical evidence corroborates the classic urban theory that physical accessibility is a fundamental prerequisite for commercial agglomeration \cite{Sevtsuk2014}. In parallel, the data confirms a significant Crowd-Mall Synergy, evidenced by the correlation ($r=0.62$) between Pedestrian Presence ($PP_i$) and Mall Spillover Vitality ($MV_i$). This finding strongly validates the Gaussian field assumption employed in our model, suggesting that large commercial anchors actively generate pedestrian flows that diffuse into the surrounding street network.

Crucially, the introduction of the Weighted Brand Ratio ($BR_i$) reveals the mechanics of quality agglomeration and provides robust validation for our semantic pipeline. To validate the economic interpretability of this VLM-derived indicator, we first examine its relationship with human mobility metrics. In street-level urban economics, sustained pedestrian footfall is a direct manifestation of revealed consumer demand. As demonstrated in Figure~\ref{FIG:10}, $BR_i$ exhibits strong positive correlations with Pedestrian Presence ($r = 0.59$) and Motor Vehicle Density ($r = 0.62$), indicating that segments with premium brand concentrations consistently draw higher consumer traffic.

To further cross-validate these findings with objective ground-truth data, we benchmarked $BR_i$ against an independent, exhaustive POI dataset. As detailed in Appendix Table~\ref{tab:external_validation}, our results reveal a striking monotonic correspondence between visual-semantic quality and physical business density. Street segments identified as "High-tier Brand Corridors" by our VLM-LLM pipeline exhibit a mean POI density 70.7\% higher (12.94) and a premium/discretionary amenity count 73.1\% higher (1.61) than "Low-tier" segments (7.58 and 0.93, respectively). This step-wise alignment with retail location theory provides compelling evidence that $BR_i$ is not merely a descriptive textual label, but a highly effective, empirically validated proxy for street-level economic quality and consumption tier.

Finally, and perhaps most intriguingly, the analysis exposes a Density-Competition Paradox. We observed a moderate positive correlation between the Closure Ratio ($CR_i$) and both Shop Density ($r=0.58$) and Brand Presence ($r=0.52$). This counter-intuitive finding reveals that high-density, high-quality commercial zones are not intrinsically stable; instead, they often experience intense competition and rapid tenant turnover. By detecting this correlation, our model successfully captures the ``high metabolic rate'' characteristic of prime locations—a phenomenon completely invisible to traditional POI-based metrics that would erroneously interpret high nominal density as absolute stability.

\subsection{Decoupling Urban Dimensions}
\begin{figure}[htbp]
    \centering
    \includegraphics[width=0.8\textwidth]{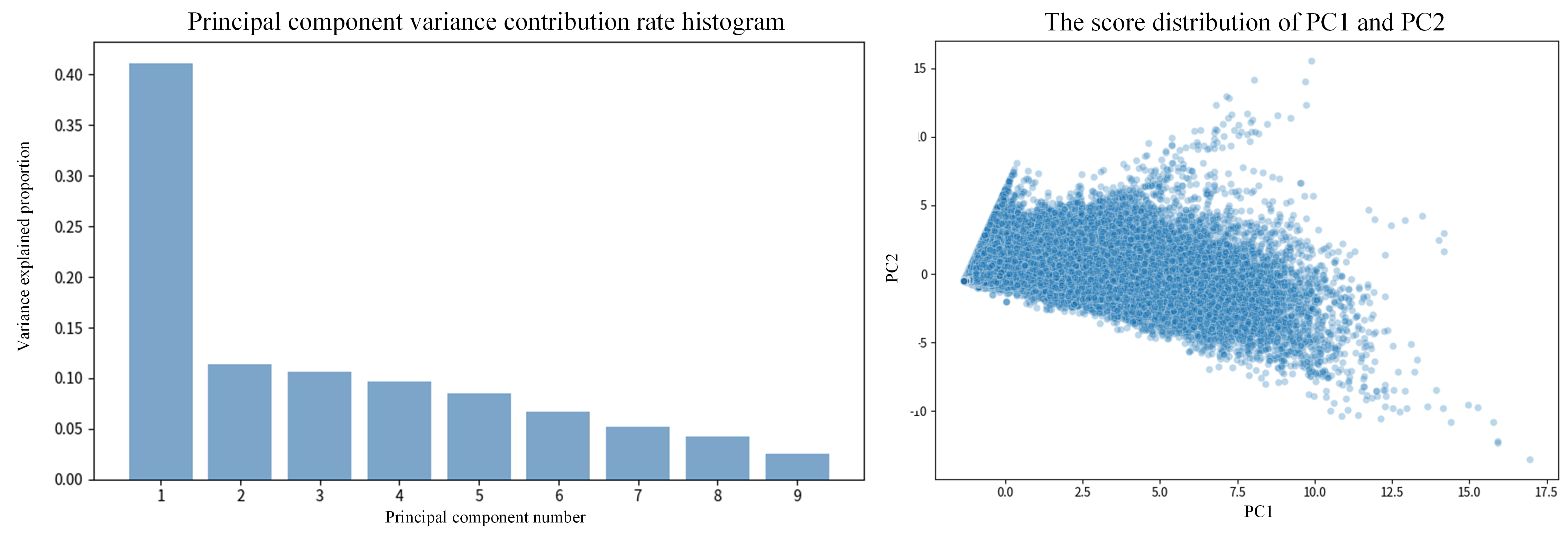}
    \caption{Principal Component Analysis results.}
    \Description{A chart visualizing the Principal Component Analysis (PCA) results and factor loadings.}
    \label{FIG:11}
\end{figure}

Principal Component Analysis (PCA) was employed to further decompose the indicator system into orthogonal latent factors. The dominant components collectively explain 72.9\% of the total variance (Figure~\ref{FIG:11}), validating the structural independence of our Diagnostic Matrix. To ensure clarity and focus on the primary economic drivers, Table~\ref{tbl4} presents the varimax-rotated factor loadings for the first four principal components.

The first component (PC1), accounting for 41.2\% of the variance, functions as the Commercial-Mobility Engine. Loading heavily on Shop Density, Motor Vehicle Density, Pedestrian Presence, and Weighted Brand Ratio, this component embodies the traditional definition of urban prosperity driven by agglomeration, accessibility, and brand quality. It represents the magnitude of street vitality.

In contrast, the second component (PC2, 11.4\%) emerges as the Ecological-Interface Trade-off Factor. Dominated by Green Coverage ($GR_i$, loading 0.745), PC2 stands in negative opposition to Shopfront Glazing Density (loading -0.375) and Shop Density (loading -0.290). This structural opposition reveals a fundamental dilemma within the historic core: streets with high ecological comfort often sacrifice commercial transparency and density. This quantitative evidence highlights a central challenge for urban renewal: how to introduce ecological interventions without dampening the visual permeability of established commercial interfaces \cite{Ye2019_LUP, Long2016_Green}.

Perhaps the most novel insight stems from the third component (PC3, 10.6\%), identified here as the Structural Recession Factor. Driven almost exclusively by the Closure Ratio ($CR_i$, loading 0.971), the emergence of decline as an orthogonal statistical dimension—distinct from low density or low brand quality—demonstrates that a dying street is fundamentally different from a quiet one. Empirically, a street can be large and dense (High PC1) yet structurally failing (High PC3). This finding confirms that without the explicit visual detection of store closures, any vitality assessment remains dimensionally incomplete.

\begin{table}[htbp]
\centering
\caption{Principal component loading matrix of key indicators (PC1 to PC4)}
\label{tbl4}
\begin{tabular}{lcccc}
\toprule
\textbf{Variables} & \textbf{PC1} & \textbf{PC2} & \textbf{PC3} & \textbf{PC4} \\
\midrule
Pedestrian Presence        & 0.411 & 0.037 & -0.094 & 0.110 \\
Motor Vehicle Density     & 0.420 & -0.019 & -0.044 & -0.088 \\
Non-Motor Vehicle Density & 0.230 & 0.328 & -0.164 & 0.800 \\
Green Coverage Ratio      & 0.149 & 0.745 & -0.081 & -0.519 \\
Weighted Brand Ratio   & 0.383 & -0.141 & 0.007 & -0.179 \\
Shop Density              & 0.437 & -0.290 & 0.026 & -0.140 \\
Shopfront Glazing Density & 0.327 & -0.375 & 0.007 & -0.056 \\
Closure Ratio             & 0.131 & 0.160 & 0.971 & 0.110 \\
Mall Spillover Vitality   & 0.344 & 0.256 & -0.106 & 0.054 \\
\bottomrule
\end{tabular}
\end{table}

\subsection{Spatial Diagnosis and Heterogeneity}
\begin{figure}[htbp]
    \centering
    \includegraphics[width=0.9\textwidth]{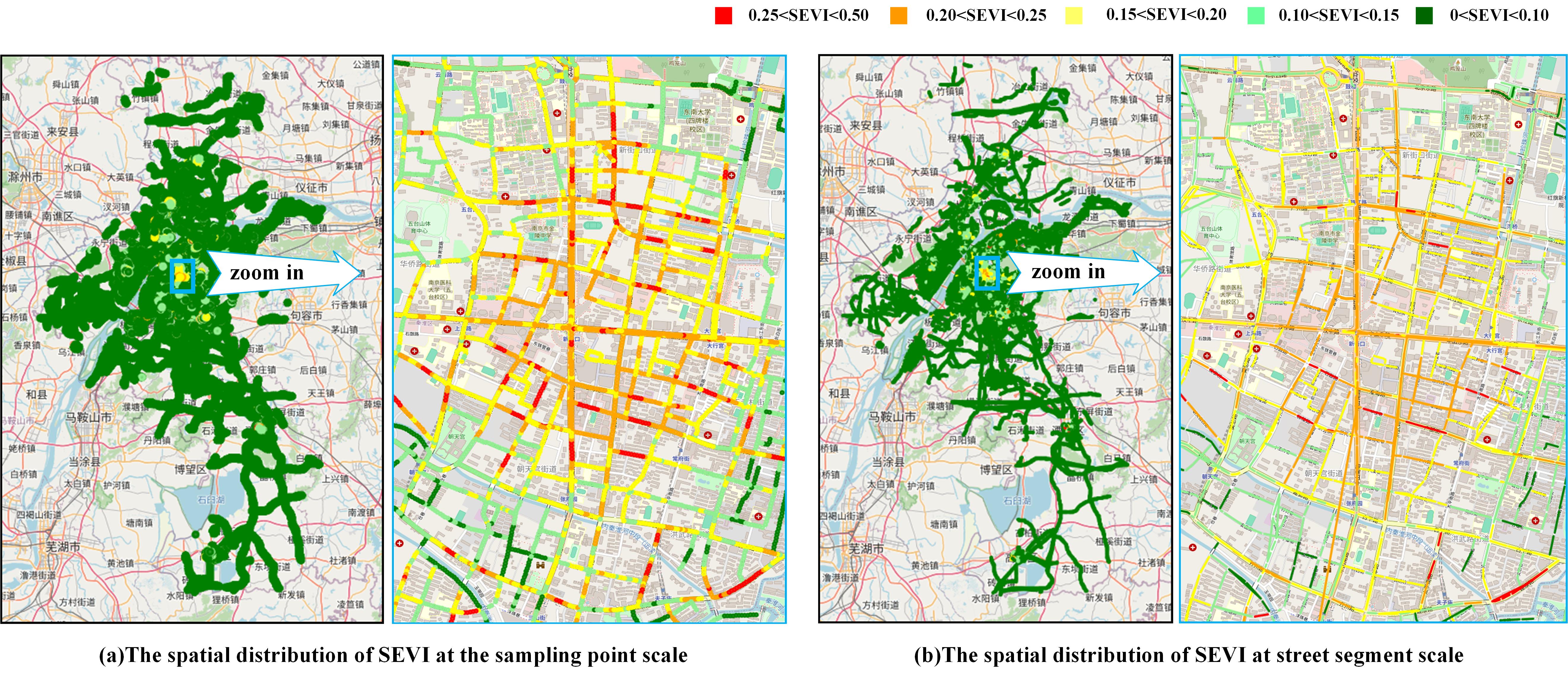}
    \caption{The distribution of SEVI.}
    \Description{A spatial distribution map showing the Street Economic Vitality Index (SEVI) across the study area.}
    \label{FIG:12}
\end{figure}

Synthesizing the multi-dimensional indicators via the EWM-TOPSIS framework, we generated a spatial distribution map of SEVI to perform a comprehensive diagnosis of Nanjing’s central area (Figure~\ref{FIG:12}). The results uncover a distinct Core-Periphery structure, nuanced by significant localized heterogeneity.

Validating the Field Effect, the Xinjiekou district—the city's commercial core—exhibits the highest SEVI scores, forming a continuous high-vitality cluster. Crucially, the vitality scores do not terminate abruptly at the boundaries of large commercial complexes; rather, they decay gradually along the connecting arteries. This observed gradient closely fits our typology-weighted Gaussian attenuation model, empirically demonstrating that the field effect of top-tier malls successfully activates the micro-circulation of the surrounding block network, physically manifesting the Crowd-Mall Synergy identified in our correlation analysis.

Beyond confirming the core, the model proves critical in distinguishing between Healthy Density and Hollow Density. Specifically, certain street segments in the historic southern districts register high traditional POI counts yet receive low SEVI scores. This discrepancy arises from the visual-semantic pipeline's dual capability: it simultaneously detects high Structural Recession (via Closure Ratio) and low Economic Quality (via Weighted Brand Ratio). By identifying streets that are nominally dense but dominated by low-tier generic shops or vacancies, the framework effectively filters out false positives in traditional data. Such diagnostic precision enables planners to prioritize these specific at-risk segments for urgent intervention, rather than misallocating resources to stable, albeit quieter, residential streets.

Furthermore, a spatial mismatch—or Green-Vitality Gap—is evident, where high-vitality zones (Red) largely coincide with low-greenery areas. This spatial pattern aligns perfectly with the Ecological-Interface Trade-off identified in PC2, highlighting the scarcity of Garden Streets in the dense core. This points to a bifurcated renewal strategy: environmental micro-intervention is prioritized for Red zones, while functional activation is required for Green/Yellow zones. Ultimately, the SEVI map functions not merely as a status visualization, but as a Computed Tomography (CT) scan of the urban tissue, revealing the hidden lesions (closures) and energy flows (spillovers) that drive the city's evolution.

\subsection{Spatiotemporal Evolution of Street Vitality Mechanisms}
While global correlation and PCA models unveil the general structural logic of street vitality, they rely on a static premise, obscuring the dynamic interactions between the built environment and human activity throughout the day. To address this, we operationalized a time-lagged Time-Sliced Geographically Weighted Regression (GWR). By assigning the static built environment features (derived from SVI and POI) to the historical baseline period ($t-1$) and the dynamic LBS crowd intensity to the subsequent observation period ($t$), we formulated the empirical model as:
\begin{equation}
    Y_{i,t} = \beta_0(u_i, v_i) + \sum_{k=1}^{n} \beta_k(u_i, v_i) X_{i,k,t-1} + \epsilon_{i,t}
\end{equation}
where $(u_i, v_i)$ represents the geographic coordinates of street segment $i$. This temporal lag structure allows us to explicitly evaluate the quasi-causal, lagged driving effects of historical street morphologies ($X_{i,t-1}$) on real-time pedestrian presence ($Y_{i,t}$), stepping beyond purely descriptive co-occurrences. Overall, this lagged GWR model demonstrates robust predictive power, achieving an average Adjusted $R^2$ of 0.66 across the observed temporal spectrum.

\begin{figure}[htbp]
    \centering
    \includegraphics[width=0.6\textwidth]{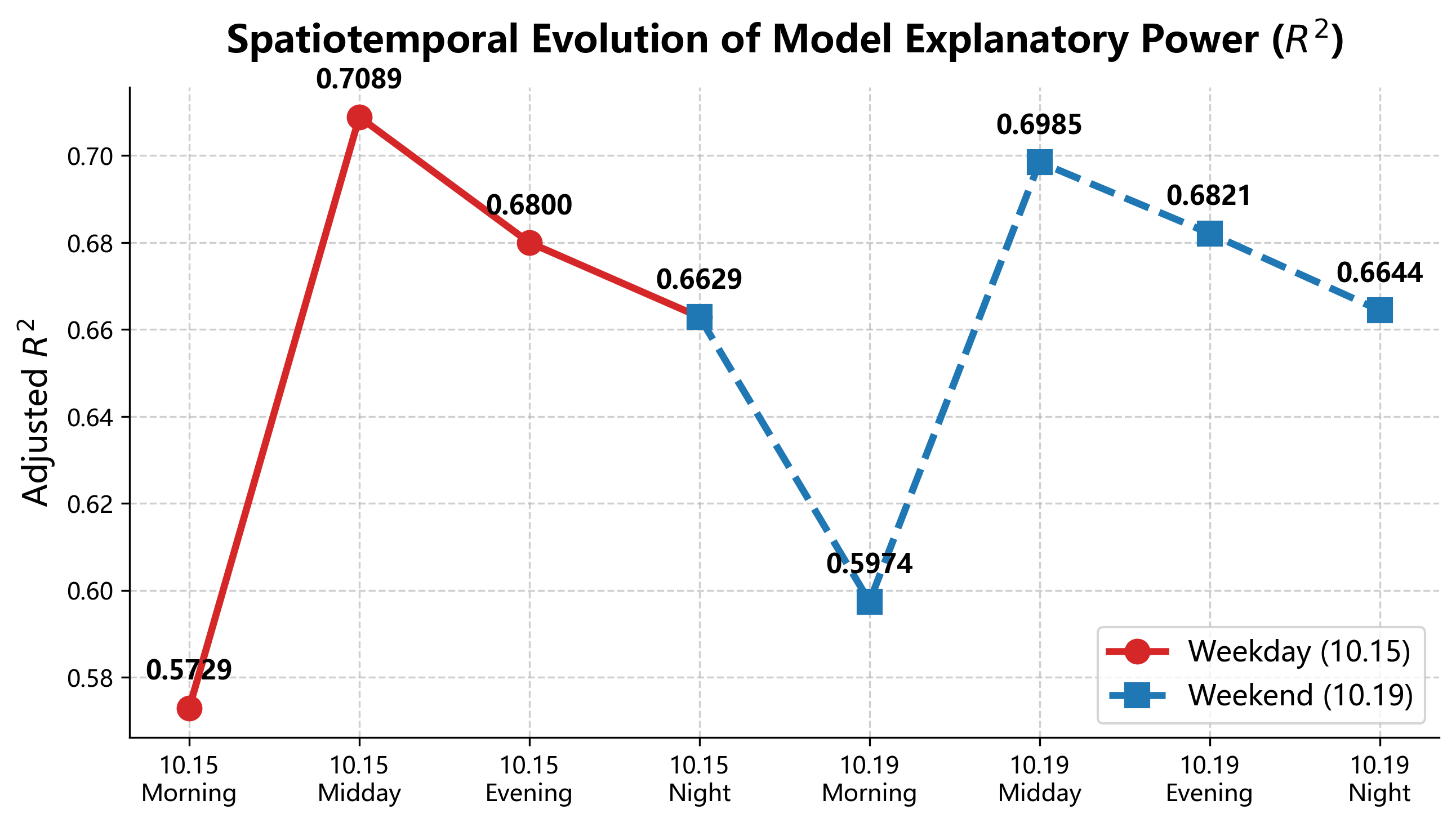}
    \caption{Spatiotemporal evolution of the Time-Sliced GWR model's explanatory power (Adjusted $R^2$).}
    \Description{A line graph showing the Adjusted R-squared values for the GWR model across eight time periods, highlighting peaks during midday and evening.}
    \label{FIG:13}
\end{figure}

The temporal evolution of the models' explanatory power (Adjusted $R^2$) reveals a profound tidal effect dictated by travel purposes (Figure~\ref{FIG:13}). During the morning peak (07:00--09:00), the model's explanatory power drops to its lowest point (Adjusted $R^2 \approx 0.57-0.59$). This suggests that morning commuting is characterized by rigid mobility; pedestrians prioritize efficiency and shortest-path routing, rendering the historical micro-scale commercial baseline ($t-1$) largely ineffective in driving subsequent foot traffic ($t$). Conversely, the model achieves its highest predictive power during the midday (11:00--14:00) and evening (17:00--19:00) periods (Adjusted $R^2$ peaking at 0.71). During these windows of discretionary activity, pedestrians have the flexibility to engage in dining and leisure, allowing the high-quality commercial interfaces of the streetscape to exert a strong, quasi-causal attractive effect on their spatial choices.

\begin{figure}[htbp]
    \centering
    \includegraphics[width=0.8\textwidth]{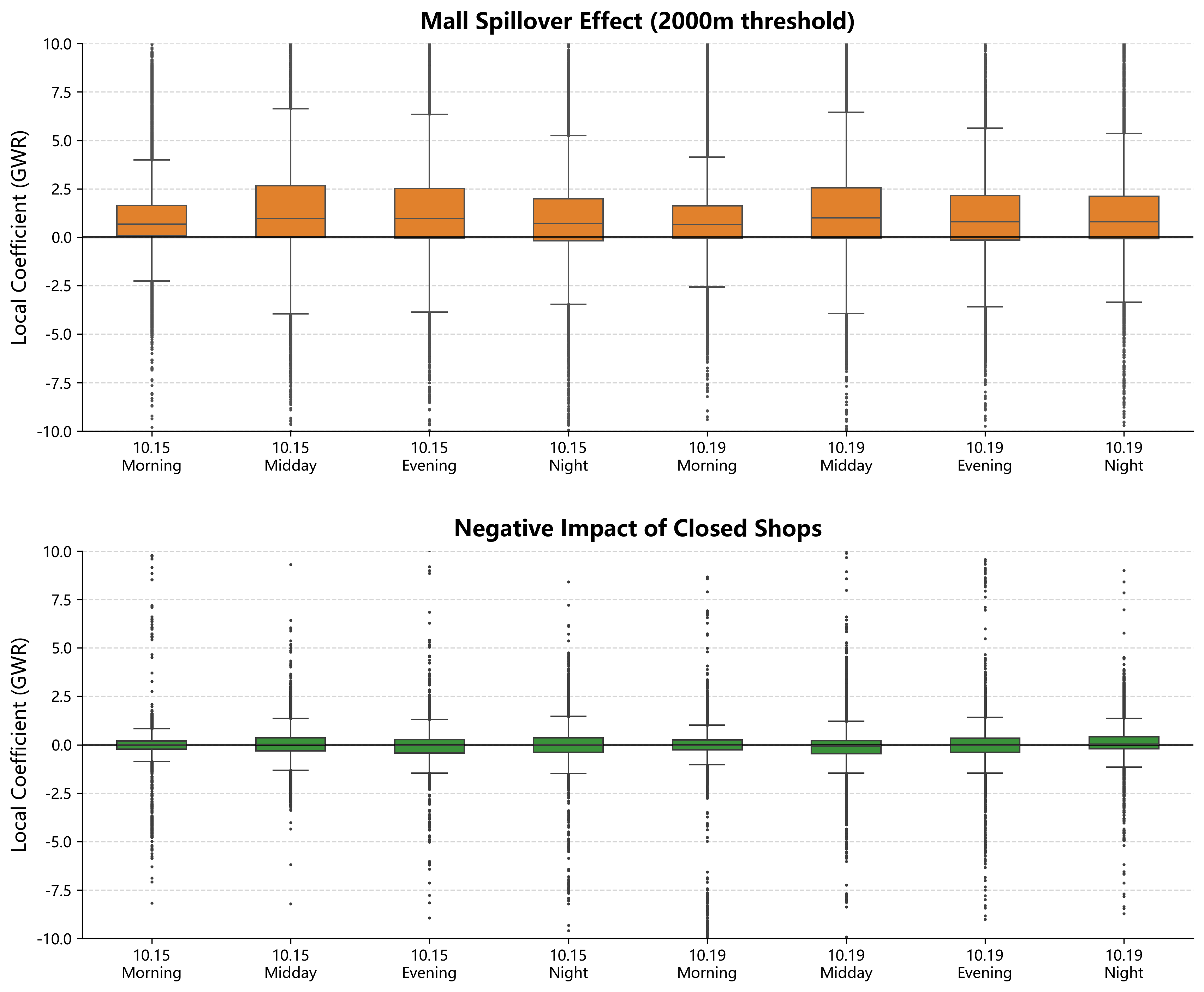}
    \caption{Temporal evolution of local GWR coefficients for Mall Spillover Vitality and Closure Ratio.}
    \Description{Two boxplots showing the distribution of local GWR coefficients across eight time periods. The top plot shows Mall Spillover Effect, and the bottom shows the negative impact of Closed Shops.}
    \label{FIG:14}
\end{figure}

Beyond global performance, analyzing the localized coefficients of specific variables uncovers the precise quasi-causal mechanics of commercial spillovers and urban decay (Figure~\ref{FIG:14}). The lagged positive influence of Mall Spillover Vitality (top panel) demonstrates significant temporal elasticity. During morning hours, the boxplot is compressed near zero, indicating minimal radiative power. However, during the midday and night economy periods, the interquartile range expands upward, confirming that shopping malls act as time-activated energy pumps, quasi-causally inducing crowd aggregation and actively exporting pedestrian flows to the surrounding 2,000-meter street network primarily during non-commuting hours. Furthermore, this spillover effect exhibits strong long-tail resilience during the weekend night economy compared to weekdays.

Equally critical is the temporal dynamic of the structural recession factor, captured by the Closure Ratio (bottom panel). While the median coefficient remains relatively stable, the lower whisker and negative outliers extend dramatically during the evening and night periods. This provides quantitative validation for Jacobs' eyes on the street theory; while closed storefronts may simply represent inactive facades during the day, at night, continuous roller shutters drastically reduce street illumination and perceived safety, exerting a severe, lagged repulsion effect that actively drives pedestrians away.

\subsection{Robustness Checks}
To ensure that our findings are not artifacts of the selected spatial constraints, we performed a series of robustness checks by re-estimating the time-lagged GWR model with alternative parameters. First, we varied the maximum spatial threshold ($D$) for the Mall Spillover Vitality ($MV_i$) from the baseline 2,000 meters to 1,000 meters and 3,000 meters. As detailed in Appendix Table A1, the model's explanatory power (Adjusted $R^2$) demonstrates remarkable stability across all thresholds. Crucially, the fundamental spatiotemporal "tidal pattern"—characterized by lowest predictive power during morning commutes and peaking during midday and evening discretionary periods—remains completely intact, confirming the robustness of the dynamic decoupling results.

Furthermore, an examination of the localized GWR coefficients corroborates the stability of our mechanistic findings. As illustrated in Appendix Figure A1, the distribution of the Mall Spillover Vitality coefficients remains structurally consistent across the 1000m, 2000m, and 3000m thresholds. In all scenarios, the median coefficients stay robustly positive, confirming the enduring positive spatial externalities of commercial anchors. More importantly, the pronounced temporal elasticity—characterized by compressed radiative power during morning commutes and significant upward expansion during midday and evening discretionary periods—is perfectly preserved regardless of the spatial constraints applied. The slight upward shift in the interquartile ranges at the 3000m threshold realistically reflects the capture of broader, long-tail radiative fields generated by top-tier mega-malls, further validating our continuous field-based assumption over rigid binary boundaries.

In addition, to address concerns regarding the functional form of spatial attenuation, we replaced the baseline Gaussian decay function with alternative Exponential and Linear decay algorithms while keeping the baseline threshold constant. As shown in Appendix Table A2, the Adjusted $R^2$ values remain remarkably stable across all three mathematical specifications. The preservation of the core spatiotemporal tidal pattern confirms that the quantified mall spillover vitality reflects an objective urban economic mechanism, rather than a mathematical artifact of a specific distance-decay formula.

Finally, to verify that the spatial diagnosis of street vitality is not mechanically driven by the chosen EWM-TOPSIS algorithm, we constructed two alternative composite indices using the Equal-Weight method ($SEVI_{EQ}$) and Principal Component Analysis ($SEVI_{PCA}$). As detailed in Appendix Table~\ref{tab:robustness_correlation}, correlation analysis reveals that the original $SEVI$ is highly consistent with both alternative indices, yielding phenomenal Spearman's rank correlation coefficients of $\rho = 0.994$ for $SEVI_{EQ}$ and $\rho = 0.977$ for $SEVI_{PCA}$ (all $p < 0.001$). This confirms that our multi-dimensional spatial diagnostic framework inherently captures the true structural variations of urban vitality, completely independent of the mathematical aggregation techniques employed.

\section{Conclusion}

This study addresses critical methodological gaps in urban vitality assessment—specifically, the semantic blindness regarding commercial quality and the spatial oversimplification of agglomeration externalities. By shifting the analytical paradigm from static POI enumeration to a visual-semantic and spatiotemporal framework, we developed the Street Economic Vitality Index (SEVI) to decode street-level economics. In doing so, we directly resolve the central economic puzzle introduced at the outset: unmasking the ``illusion of prosperity'' and providing a rigorous micro-evidence base to correct the misallocation of spatial resources during urban renewal.

\subsection*{Theoretical Contributions and Empirical Insights}
Our empirical application in the complex urban tissue of Nanjing yields four fundamental insights that refine existing urban economic theories. 

First, by transitioning from binary existence to semantic quality, we effectively corrected the observational lag in market exit. The introduction of the Closure Ratio unveiled a distinct class of pseudo-dense streets that appear thriving in nominal databases but are suffering from hidden physical recession. Crucially, the integration of the Weighted Brand Ratio—powered by our VLM-LLM pipeline—allowed us to quantify the intangible brand premium of street interfaces for the first time. We demonstrated that true economic vitality is defined not merely by the absolute density of establishments, but by the weighted agglomeration of high-tier commercial capital. 

Second, our results structurally validate the spatial externalities of commercial agglomeration. The category-weighted Gaussian spillover model successfully quantified the invisible energy field of commercial anchors, proving empirically that top-tier malls act as vitality engines with predictable, non-linear spatial decay rather than uniform buffer zones. 

Third, we revealed the profound temporal heterogeneity of vitality mechanisms. By deploying a Time-Sliced GWR against dynamic LBS data, we demonstrated that the attractive power of commercial semantics and mall spillovers is highly elastic, peaking during periods of discretionary midday and evening activity. Conversely, structural recession factors, such as shop closures, severely dampen vitality specifically during the night economy, quantitatively validating the nocturnal importance of Jacobs' ``eyes on the street'' theory regarding perceived safety and spatial friction. 

Finally, a structural tension was uncovered between commercial intensity and environmental amenity. This quantifiable prosperity-ecology trade-off highlights the inherent market failure in achieving garden-city qualities within high-density commercial cores without deliberate public intervention.

\subsection*{Implications for Urban Governance and Resource Allocation}
Beyond methodological innovation, SEVI serves as a practical diagnostic tool for precision urban governance. Unlike regional macro-metrics, our dual-scale framework enables surgical, evidence-based spatial interventions. For policymakers, the implications dictate a shift toward targeted resource allocation: high-vitality zones facing ecological deficits require targeted public goods provision (e.g., environmental micro-regeneration) to mitigate the negative externalities of extreme density. Conversely, zones characterized by hollow density require functional replacement and brand upgrading strategies rather than redundant physical infrastructure investments. Furthermore, recognizing temporal vulnerabilities allows for temporally differentiated governance; for instance, areas suffering from high nocturnal closure rates require targeted nighttime lighting and pop-up activations to offset the severe drop in pedestrian flow.

\subsection*{Limitations and Future Directions}
While this study successfully overcomes the daily snapshot limitation by integrating tidal LBS data to capture intra-day dynamics, a notable limitation remains regarding the temporal mismatch between our multi-source datasets. Due to the updating constraints of large-scale street view platforms, the SVI dataset comprises static visual slices captured in or before 2022. Although we strategically integrated this with 2023 POI data to construct a comprehensive historical baseline ($t-1$), this macro-level approach inevitably introduces a temporal resolution mismatch when regressed against the highly dynamic 2025 LBS outcomes ($t$). Specifically, this static visual baseline may fail to capture the immediate impacts of short-term, localized street micro-renewals (e.g., rapid retail turnover or sudden facade upgrades) that could instantly stimulate contemporary pedestrian flows. Future research should integrate longitudinal, time-series SVI datasets to strictly align visual variables with mobility outcomes. This would not only resolve the temporal mismatch but also enable more rigorous panel data causal inference, effectively distinguishing between short-term volatile fluctuations and permanent inter-annual economic decline. Furthermore, enriching the semantic layer by incorporating multi-sensory data (e.g., soundscapes, social media sentiment) could further deepen the human-centric resolution of the model \cite{Xu2024}.

In summary, by endowing machines with the ability to decode the semantic hierarchy and physical attrition of the street, and coupling these insights with dynamic human mobility, this research steers urban renewal away from blind, aggregate expansion toward quality-oriented, precision spatial governance.

\begin{acks}
This research is funded by National Key R\&D Program of China (2024YFC3807900).
\end{acks}

\bibliographystyle{ACM-Reference-Format}
\bibliography{sample-base}

@book{Jacobs1961,
  title={The Death and Life of Great American Cities},
  author={Jacobs, Jane},
  year={1961},
  publisher={Random House},
  address={New York}
}

@book{Gehl1987,
  title={Life Between Buildings: Using Public Space},
  author={Gehl, Jan},
  year={1987},
  publisher={Van Nostrand Reinhold},
  address={New York}
}

@article{Huang2025,
  title={Multi-Scale Street Vitality Analytics: A Comprehensive Review of Technologies, Data, and Applications},
  author={Huang, Yongming and Chen, Mingze and Zhang, Xiamengwei and Shimoda, Ryosuke and Yang, Ruochen},
  journal={Buildings},
  volume={15},
  number={21},
  pages={3987},
  year={2025},
  publisher={MDPI},
  doi={10.3390/buildings15213987}
}

@article{KOO2023,
  title={Can good microscale pedestrian streetscapes enhance the benefits of macroscale accessible urban form? An automated audit approach using {Google} street view images},
  author={Koo, Bon Woo and Guhathakurta, Subhrajit and Botchwey, Nisha and Hipp, Aaron},
  journal={Landscape and Urban Planning},
  volume={237},
  pages={104816},
  year={2023},
  publisher={Elsevier},
  doi={10.1016/j.landurbplan.2023.104816}
}

@article{Jiang2022,
  title={Street vitality and built environment features: A data-informed approach from fourteen {Chinese} cities},
  author={Jiang, Yinghong and Han, Yun and Liu, Mengyang and Ye, Yu},
  journal={Sustainable Cities and Society},
  volume={79},
  pages={103724},
  year={2022},
  publisher={Elsevier},
  doi={10.1016/j.scs.2022.103724}
}

@article{Li2022,
  title={Exploring the association between street built environment and street vitality using deep learning methods},
  author={Li, Yunqin and Yabuki, Nobuyoshi and Fukuda, Tomohiro},
  journal={Sustainable Cities and Society},
  volume={79},
  pages={103656},
  year={2022},
  publisher={Elsevier},
  doi={10.1016/j.scs.2021.103656}
}

@article{He2024,
  title={Investigating the effects of urban morphology on vitality of community life circles using machine learning and geospatial approaches},
  author={He, Sanwei and Zhang, Zhen and Yu, Shan and Xia, Chang and Tung, Chih-Lin},
  journal={Applied Geography},
  volume={167},
  pages={103287},
  year={2024},
  publisher={Elsevier},
  doi={10.1016/j.apgeog.2024.103287}
}

@article{Chen2019,
  title={Identifying urban spatial structure and urban vibrancy in highly dense cities using georeferenced social media data},
  author={Chen, Tingting and Hui, Eddie C. M. and Wu, Jiemin and Lang, Wei and Li, Xun},
  journal={Habitat International},
  volume={89},
  pages={102005},
  year={2019},
  publisher={Elsevier},
  doi={10.1016/j.habitatint.2019.102005}
}

@article{Zhang2019_SocialSensing,
  title={Social sensing from street-level imagery: A case study in learning spatio-temporal urban mobility patterns},
  author={Zhang, Fan and Zhou, Bolei and Liu, Liu and Liu, Yu and Fung, Hong H. and Lin, Hui and Ratti, Carlo},
  journal={ISPRS Journal of Photogrammetry and Remote Sensing},
  volume={153},
  pages={48--58},
  year={2019},
  publisher={Elsevier},
  doi={10.1016/j.isprsjprs.2019.04.016}
}

@article{Long2016_Green,
  title={How green are the streets? An analysis for central city of {Beijing} using {Google Street View}},
  author={Long, Ying and Liu, Lun},
  journal={Environment and Planning B: Urban Analytics and City Science},
  volume={43},
  number={6},
  pages={1118--1132},
  year={2016},
  publisher={SAGE Publications},
  doi={10.1177/0265813515600776}
}

@article{Ying2019,
  title={Does block size matter? The impact of urban design on economic vitality for {Chinese} cities},
  author={Long, Ying and Huang, C. C.},
  journal={Environment and Planning B: Urban Analytics and City Science},
  volume={46},
  number={3},
  pages={406--422},
  year={2019},
  publisher={SAGE Publications},
  doi={10.1177/2399808317715640}
}

@article{Lan2020,
  title={How do population inflow and social infrastructure affect urban vitality? Evidence from {Shanghai}, {China}},
  author={Lan, Feng and Gong, Xiaoqing and Da, Haizhi and Wen, Haizhen},
  journal={Cities},
  volume={100},
  pages={102659},
  year={2020},
  publisher={Elsevier},
  doi={10.1016/j.cities.2020.102659}
}

@article{Xu2024,
  title={Assessing urban street vitality through visual and auditory perception: A case study of historic urban area in {Guangzhou}, {China}},
  author={Xu, Yuhan and Ma, Xiaosu},
  journal={International Review for Spatial Planning and Sustainable Development},
  volume={12},
  number={4},
  pages={57--76},
  year={2024},
  publisher={SPSD Press},
  doi={10.14246/irspsd.12.4_57}
}

@article{Ling2025,
  title={Unveiling the hidden vitality of block-structured neighborhoods through a multimodal urban perception and ensemble learning framework},
  author={Ling, Zhenxiang and Meng, Xianxin and Chen, Yingbiao and Qian, Qinglan and Kuang, Junyu and Shi, Xianghua and Yang, Yifan and Chen, Wentao and Zheng, Zihao and Wu, Zhifeng},
  journal={International Journal of Digital Earth},
  volume={18},
  number={1},
  pages={2545581},
  year={2025},
  publisher={Taylor \& Francis},
  doi={10.1080/17538947.2025.2545581}
}

@article{Wang2022_JUM,
  title={The influence of street network configuration on street vitality in {Singapore}},
  author={Wang, X. and Yuen, B.},
  journal={Journal of Urban Management},
  volume={11},
  number={2},
  pages={202--215},
  year={2022},
  publisher={Elsevier},
  doi={10.1016/j.jum.2022.01.002}
}

@article{Liao2025,
  title={Exploring the causal relationship between campus walkability and affective walking experience: Evidence from 7 major tertiary education campuses in {China}},
  author={Liao, Bojing and Zhu, Jie},
  journal={Journal of Urban Management},
  volume={14},
  number={3},
  pages={657--674},
  year={2025},
  publisher={Elsevier},
  doi={10.1016/j.jum.2025.01.005}
}

@article{Sevtsuk2014,
  title={Location and agglomeration of retail and food services: The case of {Somerville}, {MA}},
  author={Sevtsuk, Andres},
  journal={Urban Studies},
  volume={51},
  number={16},
  pages={3745--3764}, 
  year={2014},
  publisher={SAGE Publications},
  doi={10.1177/0042098013516523}
}

@article{Yang2024,
  title={Understanding urban vitality from the economic and human activities perspective: A case study of {Chongqing}, {China}},
  author={Yang, Fiona Fan and Lin, Geng and Lei, Yubing and Wang, Ying and Yi, Zheng},
  journal={Chinese Geographical Science},
  volume={34},
  number={1},
  pages={52--66},
  year={2024},
  publisher={Springer},
  doi={10.1007/s11769-023-1402-2}
}

@article{Chen2022,
  title={Evaluating urban vitality based on geospatial big data in {Xiamen Island}, {China}},
  author={Chen, Shili and Lang, Wei and Li, Xun},
  journal={Sage Open},
  volume={12},
  number={4},
  year={2022},
  publisher={SAGE Publications},
  doi={10.1177/21582440221134519}
}

@misc{NanjingPlan2024,
  author={{Nanjing Municipal Bureau of Planning and Natural Resources}},
  title={{Territorial and Spatial Master Plan of Nanjing (2021--2035)}},
  year={2024},
  url={https://ghj.nanjing.gov.cn/ghbz/ztgh/202410/t20241024_4992742.html},
  note={Accessed: 2025-12-22}
}

@inproceedings{Redmon2016_YOLO,
  title={You only look once: Unified, real-time object detection},
  author={Redmon, Joseph and Divvala, Santosh and Girshick, Ross and Farhadi, Ali},
  booktitle={Proceedings of the IEEE Conference on Computer Vision and Pattern Recognition},
  pages={779--788},
  year={2016},
  publisher={IEEE},
  doi={10.1109/CVPR.2016.91}
}

@inproceedings{Lin2014,
  title={{Microsoft} {COCO}: Common objects in context},
  author={Lin, Tsung-Yi and Maire, Michael and Belongie, Serge and Hays, James and Perona, Pietro and Ramanan, Deva and Doll{\'a}r, Piotr and Zitnick, C. Lawrence},
  booktitle={European Conference on Computer Vision},
  pages={740--755},
  year={2014},
  publisher={Springer},
  doi={10.1007/978-3-319-10602-1_48}
}

@misc{Wada2021,
  author={Wada, Kentaro},
  title={{Labelme}: Image Polygonal Annotation with {Python}},
  year={2021},
  url={https://github.com/wkentaro/labelme}
}

@misc{Jocher2023,
  author={Jocher, Glenn and Chaurasia, Ayush and Laughing and others},
  title={{Ultralytics} {YOLOv8}},
  year={2023},
  publisher={Zenodo},
  doi={10.5281/zenodo.7841070}
}

@article{Zhu2020_EWM,
  title={Effectiveness of Entropy Weight Method in Decision-Making},
  author={Zhu, Yuxin and Tian, Dazuo and Yan, Feng},
  journal={Mathematical Problems in Engineering},
  volume={2020},
  pages={1--5},
  year={2020},
  publisher={Hindawi},
  doi={10.1155/2020/3564835}
}

@book{Hwang1981,
  title={Multiple Attribute Decision Making: Methods and Applications},
  author={Hwang, Ching-Lai and Yoon, Kwangsun},
  year={1981},
  publisher={Springer-Verlag},
  address={New York},
  doi={10.1007/978-3-642-48318-9}
}

@article{Yue2017,
  title={Measurements of {POI}-based mixed use and their relationships with neighbourhood vibrancy},
  author={Yue, Yang and Zhuang, Yan and Yeh, Anthony G. O. and Xie, Jinyu and Ma, Chengling and Li, Qingquan},
  journal={International Journal of Geographical Information Science},
  volume={31},
  number={4},
  pages={658--675},
  year={2017},
  publisher={Taylor \& Francis},
  doi={10.1080/13658816.2016.1220561}
}

@article{Mehta2009,
  title={Look closely: The hedonic value of walkable streets},
  author={Mehta, Vikas},
  journal={Journal of Urban Design},
  volume={14},
  number={2},
  pages={213--241},
  year={2009},
  publisher={Taylor \& Francis},
  doi={10.1080/13574800802670929}
}

@article{Ye2019_LUP,
  title={Measuring daily accessed street greenery: A human-scale approach for informing better urban planning},
  author={Ye, Yu and Richards, Daniel and Lu, Yi and Song, Xiaoqing and Zhuang, Yan and Zeng, Wei and Zhong, Teng},
  journal={Landscape and Urban Planning},
  volume={191},
  pages={103434},
  year={2019},
  publisher={Elsevier},
  doi={10.1016/j.landurbplan.2019.103434}
}

@article{Li2022_IJGI,
  title={Multidimensional urban vitality on streets: Spatial patterns and influence factor identification using multisource urban data},
  author={Li, Qian and Cui, Caihui and Liu, Feng and Wu, Qirui and Run, Yadi and Han, Zhigang},
  journal={ISPRS International Journal of Geo-Information},
  volume={11},
  number={1},
  pages={2},
  year={2022},
  publisher={MDPI},
  doi={10.3390/ijgi11010002}
}

@article{Zhang2021_EPB,
  title={How can the urban landscape affect urban vitality at the street block level? A case study of 15 metropolises in {China}},
  author={Zhang, Anqi and Li, Weifeng and Wu, Jiayu and Lin, Jian and Chu, Jianqun and Xia, Chang},
  journal={Environment and Planning B: Urban Analytics and City Science},
  volume={48},
  number={5},
  pages={1245--1262},
  year={2021},
  publisher={SAGE Publications},
  doi={10.1177/2399808320924425}
}

@inproceedings{UrbanVLP2025,
  title={UrbanVLP: Multi-Granularity Vision-Language Pretraining for Urban Socioeconomic Indicator Prediction},
  author={Hao, Xixuan and Chen, Wei and Yan, Yibo and Zhong, Siru and Wang, Kun and Wen, Qingsong and Liang, Yuxuan},
  booktitle={Proceedings of the AAAI Conference on Artificial Intelligence},
  volume={39},
  number={27},
  pages={28061--28069},
  year={2025}
}

@article{MINGLE2025,
  title={MINGLE: VLMs for Semantically Complex Region Detection in Urban Scenes},
  author={Liu, Liu and Kudaeva, Alexandra and Cipriano, Marco and Al Ghannam, Fatimeh and Tan, Freya and de Melo, Gerard and Sevtsuk, Andres},
  journal={arXiv preprint arXiv:2509.13484},
  year={2025}
}

@article{VLMReview2025,
  title={Vision language model ({VLM})-enabled street view analytics: a systematic literature review},
  author={Peng, Ziyu and Lu, Weisheng and An, Hongda and Xia, Xianhua and Zhang, Yi and Xue, Fan and Chen, Junjie},
  journal={Engineering, Construction and Architectural Management},
  year={2025},
  publisher={Emerald Publishing Limited},
  doi={10.1108/ECAM-07-2025-1133}
}

\appendix
\setcounter{table}{0}
\renewcommand{\thetable}{A\arabic{table}}
\setcounter{figure}{0}
\renewcommand{\thefigure}{A\arabic{figure}}

\subsection{Robustness Checks Results}

To validate the structural consistency of the Mall Spillover Vitality ($MV_i$), we performed robustness checks by varying the maximum spatial threshold ($D$) from the baseline 2,000 meters to 1,000 meters and 3,000 meters. 

\begin{figure}[htbp]
  \centering
  \includegraphics[width=\textwidth]{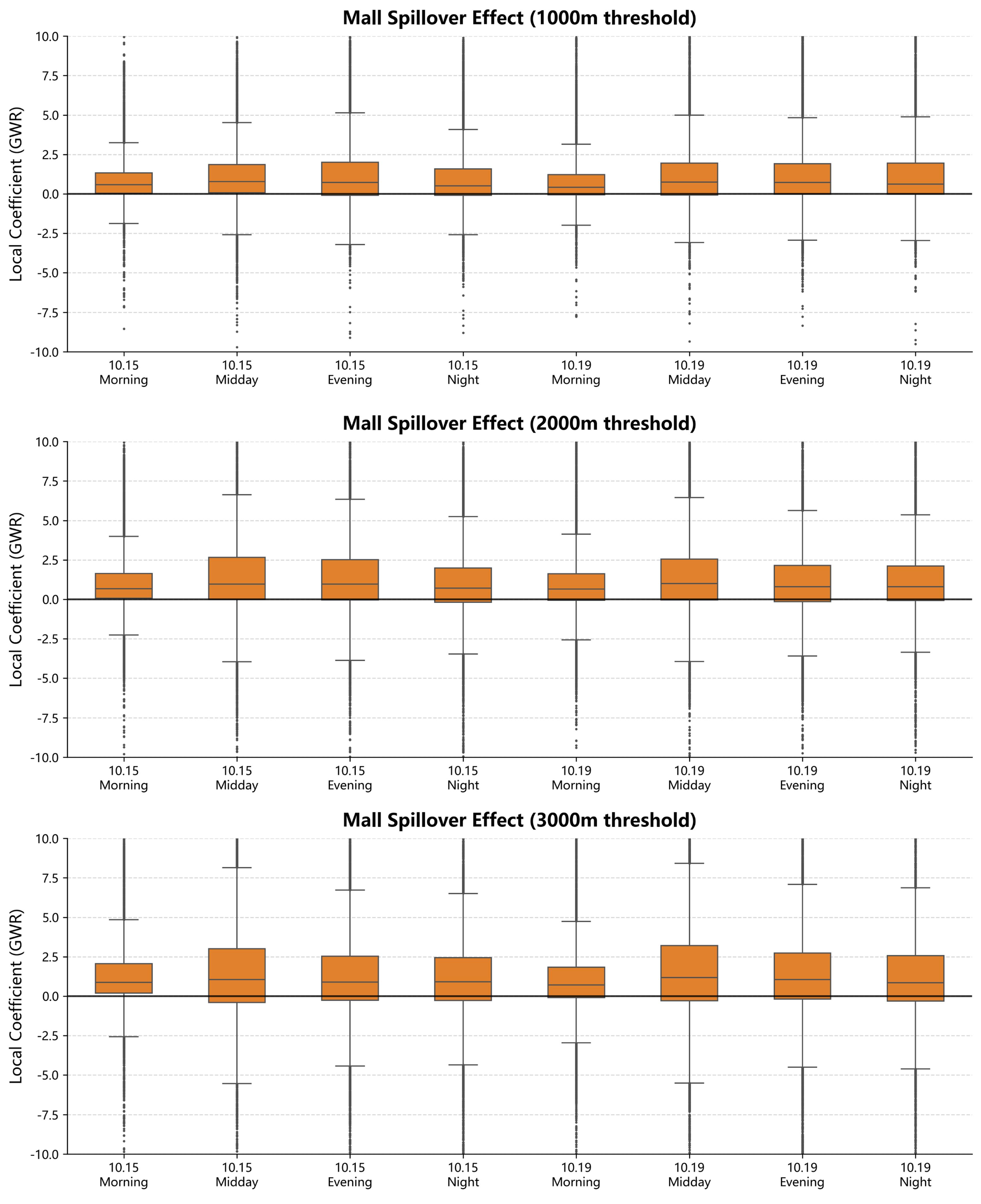}
  \caption{Temporal evolution of local GWR coefficients for Mall Spillover Vitality under alternative spatial thresholds (1000m, 2000m, and 3000m). The consistent positive medians and expanding interquartile ranges during non-commuting hours confirm the structural robustness of the field-based spillover effect.}
  \label{fig:robustness_boxplots}
\end{figure}

Table~\ref{tab:robustness_r2} presents the Adjusted $R^2$ values of the time-lagged GWR models under these alternative thresholds. The results demonstrate remarkable stability across all temporal periods, perfectly preserving the spatiotemporal "tidal pattern" characterized by lower predictive power during morning commutes and peaking during discretionary periods. Figure~\ref{fig:robustness_boxplots} further illustrates the localized GWR coefficients for the 1000m, 2000m, and 3000m thresholds. The persistent positive medians and coherent temporal elasticity confirm that the quasi-causal spatial externalities of commercial anchors are highly robust and do not rely on a specifically rigid spatial boundary.

\begin{table}[htbp]
  \centering
  \caption{Comparison of Time-Lagged GWR Model Explanatory Power (Adjusted $R^2$) under Varying Spatial Thresholds}
  \label{tab:robustness_r2}
  \begin{tabular}{l c c c}
    \toprule
    \textbf{Time Period} & \textbf{1000m} & \textbf{2000m (Baseline)} & \textbf{3000m} \\
    \midrule
    Weekday (10.15) Morning & 0.5745 & 0.5729 & 0.5743 \\
    Weekday (10.15) Midday  & 0.7125 & 0.7089 & 0.7164 \\
    Weekday (10.15) Evening & 0.6840 & 0.6800 & 0.6808 \\
    Weekday (10.15) Night   & 0.6609 & 0.6629 & 0.6620 \\
    \addlinespace
    Weekend (10.19) Morning & 0.5945 & 0.5974 & 0.5942 \\
    Weekend (10.19) Midday  & 0.7006 & 0.6985 & 0.7030 \\
    Weekend (10.19) Evening & 0.6807 & 0.6821 & 0.6814 \\
    Weekend (10.19) Night   & 0.6674 & 0.6644 & 0.6673 \\
    \bottomrule
  \end{tabular}
\end{table}

\begin{table}[htbp]
  \centering
  \caption{Comparison of Time-Lagged GWR Model Explanatory Power (Adjusted $R^2$) under Alternative Distance-Decay Functions}
  \label{tab:robustness_decay_functions}
  \begin{tabular}{l c c c}
    \toprule
    \textbf{Time Period} & \textbf{Gaussian Decay (Baseline)} & \textbf{Exponential Decay} & \textbf{Linear Decay} \\
    \midrule
    Weekday (10.15) Morning & 0.5729 & 0.5734 & 0.5738 \\
    Weekday (10.15) Midday  & 0.7089 & 0.7164 & 0.7121 \\
    Weekday (10.15) Evening & 0.6800 & 0.6808 & 0.6838 \\
    Weekday (10.15) Night   & 0.6629 & 0.6623 & 0.6695 \\
    \addlinespace
    Weekend (10.19) Morning & 0.5974 & 0.5970 & 0.5973 \\
    Weekend (10.19) Midday  & 0.6985 & 0.7041 & 0.7049 \\
    Weekend (10.19) Evening & 0.6821 & 0.6833 & 0.6866 \\
    Weekend (10.19) Night   & 0.6644 & 0.6653 & 0.6664 \\
    \bottomrule
  \end{tabular}
\end{table}

\begin{table}[htbp]
  \centering
  \caption{Spearman Rank Correlation Matrix of Alternative SEVI Constructs}
  \label{tab:robustness_correlation}
  \begin{tabular}{l c c c}
    \toprule
    \textbf{Indices} & \textbf{SEVI$_{EWM-TOPSIS}$} & \textbf{SEVI$_{EQ}$} & \textbf{SEVI$_{PCA}$} \\
    \midrule
    \textbf{SEVI$_{EWM-TOPSIS}$} & 1.000 & 0.994$^{***}$ & 0.977$^{***}$ \\
    \textbf{SEVI$_{EQ}$}         & 0.994$^{***}$ & 1.000 & 0.984$^{***}$ \\
    \textbf{SEVI$_{PCA}$}        & 0.977$^{***}$ & 0.984$^{***}$ & 1.000 \\
    \bottomrule
  \end{tabular}
  
  \vspace{1.5mm}
  {\small \textit{Note:} *** denotes statistical significance at the $p < 0.001$ level.}
\end{table}

\section{External Validation with Multi-Source POI Data}

To ensure the robustness and economic interpretability of the VLM-derived Weighted Brand Ratio ($BR_i$), we conducted a comprehensive external validation using an independent, large-scale Point-of-Interest (POI) dataset of Nanjing. The dataset comprises 249,257 validated commercial establishments, including Catering Services ($n = 71,173$), Shopping Services ($n = 100,878$), Sports and Leisure ($n = 12,370$), and Life Services ($n = 64,836$), providing a high-resolution representation of urban commercial activity.

\subsection{Spatial Matching and Buffering}

To align the semantic measure $BR_i$ with physical commercial patterns, we performed a spatial join between 280,528 street-level sampling points and the POI dataset. A 50-meter Euclidean buffer was applied to each sampling point to capture the immediate commercial catchment area. To ensure computational efficiency for this large-scale spatial query, we implemented a \textit{cKDTree} (Coherent K-Dimensional Tree) algorithm for neighborhood search, achieving $O(\log n)$ query complexity.

\subsection{Definition of Commercial Activity and Filtering}

To focus on meaningful commercial environments and reduce spatial noise (e.g., isolated POIs in predominantly residential areas), we retained only sampling points with at least one POI within the 50-meter radius. This filtering step resulted in 63,821 active commercial segments, forming the basis for subsequent validation analysis.

\subsection{Statistical Validation}

Given the skewed distribution of $BR_i$, we employed Spearman’s rank correlation to assess the monotonic relationship between the VLM-derived brand metric and observed commercial intensity. In addition, a tertile-based stratification (Low, Mid, High tiers) was applied to examine differences in POI density and composition across varying levels of $BR_i$.

The results (Table~\ref{tab:external_validation}) demonstrate that $BR_i$ strongly correlates with both the density and hierarchical composition of commercial establishments, confirming its effectiveness as a proxy for the quality and agglomeration of urban economic activity. All observed relationships are statistically significant ($p < 0.001$).

\begin{table}[htbp]
  \centering
  \caption{External Validation of $BR_i$ against Physical Business Density}
  \label{tab:external_validation}
  \begin{tabular}{l c c}
    \toprule
    \textbf{$BR_i$ Intensity Tier} & \textbf{Mean Total POI Count} & \textbf{Mean Premium/Discretionary POI Count} \\
    \midrule
    Low-tier Brand Segments  & 7.58  & 0.93 \\
    Mid-tier Brand Segments  & 10.00 & 1.24 \\
    High-tier Brand Segments & 12.94 & 1.61 \\
    \midrule
    \textbf{Growth (High vs. Low)} & \textbf{+70.7\%} & \textbf{+73.1\%} \\
    \bottomrule
  \end{tabular}
  \vspace{1.5mm}
  {\small \textit{Note:} $BR_i$ tiers are based on tertile splitting of the VLM-derived Weighted Brand Ratio. All differences between groups are statistically significant ($p < 0.001$, Kruskal-Wallis test).}
\end{table}

\end{document}